\documentclass[reprint,preprintnumbers,superscriptaddress,amsmath,aps,prd,linenumbers]{revtex4-2}

\usepackage{color}
\usepackage{enumitem}
\usepackage{lscape}
\usepackage{rotating}
\usepackage{amsmath}
\usepackage{graphicx} 
\usepackage{epstopdf} 
\usepackage{dcolumn} 
\usepackage{xcolor} 
\usepackage{amsmath,amssymb} 
\usepackage[utf8]{inputenc} 
\usepackage[english]{babel} 
\usepackage{blindtext} 
\usepackage{ifthen} 
\usepackage{orcidlink} 
\usepackage{hyperref} 

\graphicspath{{figures/}} 

\newboolean{articletitles}
\setboolean{articletitles}{true} 

\makeatletter
\patchcmd{\@outputpage@head}{\@ifx{\LS@rot\@undefined}{}{\LS@rot}}{}{}{}
\makeatother

\input{belle2-symbols}
\newcommand*\patchAmsMathEnvironmentForLineno[1]{%
  \expandafter\let\csname old#1\expandafter\endcsname\csname #1\endcsname
  \expandafter\let\csname oldend#1\expandafter\endcsname\csname end#1\endcsname
  \renewenvironment{#1}%
     {\linenomath\csname old#1\endcsname}%
     {\csname oldend#1\endcsname\endlinenomath}}%
\newcommand*\patchBothAmsMathEnvironmentsForLineno[1]{%
  \patchAmsMathEnvironmentForLineno{#1}%
  \patchAmsMathEnvironmentForLineno{#1*}}%
\AtBeginDocument{%
\patchBothAmsMathEnvironmentsForLineno{equation}%
\patchBothAmsMathEnvironmentsForLineno{align}%
\patchBothAmsMathEnvironmentsForLineno{flalign}%
\patchBothAmsMathEnvironmentsForLineno{alignat}%
\patchBothAmsMathEnvironmentsForLineno{gather}%
\patchBothAmsMathEnvironmentsForLineno{multline}%
}

\newcommand{\kekpreprint}{2024-23}  
\newcommand{\biipreprint}{2024-024}  
\begin{document}
\nolinenumbers

\begin{flushright}
    ~\\
    ~\\
    KEK preprint \kekpreprint\\
    Belle II preprint \biipreprint\\
    \end{flushright}
\title{Measurement of the branching fraction, polarization, and time-dependent {\it{ CP} } asymmetry in $B^0 \to \rho^+\rho^-$ decays and constraint on the CKM angle $\phi_2$}

  \author{I.~Adachi\,\orcidlink{0000-0003-2287-0173}} 
  \author{L.~Aggarwal\,\orcidlink{0000-0002-0909-7537}} 
  \author{H.~Ahmed\,\orcidlink{0000-0003-3976-7498}} 
  \author{N.~Akopov\,\orcidlink{0000-0002-4425-2096}} 
  \author{M.~Alhakami\,\orcidlink{0000-0002-2234-8628}} 
  \author{A.~Aloisio\,\orcidlink{0000-0002-3883-6693}} 
  \author{N.~Althubiti\,\orcidlink{0000-0003-1513-0409}} 
  \author{N.~Anh~Ky\,\orcidlink{0000-0003-0471-197X}} 
  \author{D.~M.~Asner\,\orcidlink{0000-0002-1586-5790}} 
  \author{H.~Atmacan\,\orcidlink{0000-0003-2435-501X}} 
% ...
\collaboration{The Belle II Collaboration}
\begin{abstract}
We present a measurement of the branching fraction and fraction of longitudinal polarization 
of $B^0 \to \rho^+ \rho^-$ decays, which have two $\pi^0$'s in the final state. We also measure time-dependent {\it CP} violation parameters for decays into
longitudinally polarized $\rho^+ \rho^-$ pairs.
This analysis is based on a data sample containing $(387\pm6) \times 10^6$ $\Upsilon(4S)$ mesons collected with the Belle~II detector at the SuperKEKB asymmetric-energy $e^+e^-$ collider in 2019-2022.
We obtain $\mathcal{B}(B^0\to\rho^+\rho^-) = (2.89 ^{+0.23}_{-0.22} {}^{+0.29}_{-0.27}) \times 10^{-5}, f_{L} = 0.921 ^{+0.024}_{-0.025} {}^{+0.017}_{-0.015}$, \mbox{$S = -0.26\pm0.19\pm0.08$}, and $C = -0.02\pm0.12^{+0.06}_{-0.05}$, 
where the first uncertainties are statistical and the second are systematic.
We use these results to perform an isospin analysis to constrain the CKM angle $\phi_2$ and obtain two solutions; the result consistent with other Standard Model constraints is $\phi_2 = (92.6^{+4.5}_{-4.7})^\circ$.

\end{abstract}
\maketitle
\renewcommand{\figurename}{Fig.}
\section{Introduction}
Charge-parity ({\it{CP}}) violation in the Standard Model (SM) is described by a single irreducible complex phase in the Cabibbo-Kobayashi-Maskawa (CKM) quark-mixing matrix~\cite{Cabibbo:1963yz,Kobayashi:1973fv}.
Measurements of {\it CP} asymmetries, mixing frequencies, and branching fractions~(${\mathcal{B}}$) of $B$ hadron decays constrain the angles and sides of the CKM unitarity triangle~(UT)~\cite{UTfit,Charles_2005}.
Although current measurements are consistent with the CKM picture of the SM, the precision still allows for  
${\mathcal{O}}$(10)\% non-SM contributions to the $B^0$-$\Bzb$ mixing amplitudes~\cite{UTfit:2007eik,Altmannshofer:2013lfa,Buras:2013ooa,Tanimoto:2014eva,Charles:2015gya,PhysRevD.102.056023}. 
Hence, further improvement of experimental and theoretical knowledge of the UT can help to identify or constrain non-SM physics.

The angle $\phi_2$~\footnote{The angle $\phi_2$ is also known as $\alpha$.}, which is defined in terms of CKM matrix elements as ${\rm{arg}} (-V_{td}V_{tb}^*/V_{ud}V_{ub}^*)$, is the least known angle of the UT; the current world average is $\phi_2 = (84.1^{+4.5}_{-3.8})^{\circ}$~\cite{ParticleDataGroup:2024xxx}.
We can determine the angle $\phi_2$ by measuring the time-dependent {\it CP} asymmetry between $B^0$ and $\Bzb$ decays proceeding via $b \to \uubar d$ transitions.

In $e^+e^-$ collisions at the $\Upsilon(4S)$ resonance, a quantum-entangled $B^0\Bzb$ pair is produced via $e^+e^- \to \Upsilon(4S) \to B^0\Bzb$. 
The probability to observe a $B$ meson decaying into a {\it CP} eigenstate~($B_{\it CP}$) at proper time $t_{\it{CP}}$ while the other $B$ meson~($B_{\rm tag}$) decays with flavor $q$ ($q=+1$ for $B^0$ and $q=-1$ for $\Bzb$) at proper time $t_{\rm{tag}}$ is given by
\begin{eqnarray}
P(\Delta t, q) = \frac{e^{{-|\Delta t|}/\tau_{B^0}}}{4\tau_{B^0}} \Bigl\{ 1 + q\bigl[ S\sin (\Delta m_d \Delta t) \nonumber \\
- C\cos (\Delta m_d \Delta t) \bigr]   \Bigr\},
\end{eqnarray}
where $\Delta t \equiv t_{\it{CP}} - t_{\rm{tag}}$, $\tau_{B^0}$ is the lifetime of the $B^0$ meson, and $\Delta m_d$ is the mass difference between the two $B^0$ mass eigenstates~{\cite{ParticleDataGroup:2024xxx}\footnote{We use a system of units in which $c=\hbar=1$.}. Here, $S$ and $C$~\footnote{The {\it CP} violation parameter $C$ is also defined as $A=-C$.} }
are mixing-induced and direct {\it CP} violation parameters~\footnote{
We assume no {\it CP} violation in $B^0$-$\Bzb$ mixing, which is an excellent approximation.
}, respectively.

In $B$ meson decays, the tree level $b \to \uubar d$ process is the leading amplitude in $B \to \pi \pi$, $B \to \rho \pi$, $B \to \rho \rho$ and $B \to a_1 \pi$ decays. This amplitude has a weak phase of $\phi_2$, including the phase from $B^0$-$\Bzb$ mixing. In addition, a $b\to d$ loop amplitude contributes to these decays at a sub-leading level.
This additional amplitude has a different weak phase from the tree amplitude and thus shifts the value of $S$ from $\sin{(2\phi_2)}$ to $\sqrt{1-C^2}\sin{(2\phi_2 + \Delta\phi_2)}$.
To estimate the effect of the loop amplitude and the shift $\Delta \phi_2$, an isospin analysis using the branching fractions and direct {\it CP} asymmetries in those charmless decays is used~\cite{Gronau:1990ka}. Such an analysis shows that $B^0 \to \rho^+ \rho^-$~\footnote{Throughout this paper, the inclusion of the charge conjugate decay mode is implied unless otherwise specified.} has a small contribution from the loop amplitude~\cite{BaBar:2007cku,Belle:2015xfb,Belle:2003lsm,BaBar:2006sla,BaBar:2008xku,Belle:2012ayg,LHCb:2015zxm} and gives a more stringent constraint on $\phi_2$~\cite{Charles:2017evz} than those resulting from $B\to\pi\pi$, $B\to\rho\pi$, and $B\to a_1\pi$ decays.
{
\tabcolsep = 10pt

\begin{table*}[htb]
  \caption[]{Recent precise measurements and world averages of the branching fraction, fraction of longitudinal polarization, and {\it CP} violation parameters for $B \to \rho^+ \rho^-$ decays. Note that both the Belle and \babar experiments assume equal production of neutral and charged $B$ meson pairs from $\Upsilon(4S)$ decay for their branching fraction measurements.
  }
  \label{tab:prev}
  \centering
  \begin{tabular*}{\linewidth}{@{\extracolsep{\fill}}lcccc}
  \hline \hline
  Exp. & ${\mathcal{B}}$ [$10^{-5}$] & $f_{L}$ & $S$ & $C$ \\
  \hline
  Belle~\cite{Belle:2015xfb}     & $2.83 \pm 0.15 \pm 0.15$           & $0.988 \pm 0.012 \pm 0.023$ & $-0.13 \pm 0.15 \pm 0.05$ & $0.00 \pm 0.10 \pm 0.06$ \\
  \babar~\cite{BaBar:2007cku}     & $2.55 \pm 0.21 {}^{+0.36}_{-0.39}$ & $0.992 \pm 0.024 {}^{+0.026}_{-0.013}$ & $-0.17 \pm 0.20^{+0.05}_{-0.06}$ & $0.01 \pm 0.15 \pm 0.06$\\
  \hline
  PDG~\cite{ParticleDataGroup:2024xxx}       & $2.77 \pm 0.19$                    & $0.990^{+0.021}_{-0.019}$ & $-0.14 \pm 0.13$ & $0.00 \pm 0.09$\\
  \hline \hline
  \end{tabular*}
\end{table*}
}

Since $B^0 \to \rho^+ \rho^-$ is the decay of a pseudo-scalar to two vector mesons, there are three helicity states of the $\rho$ meson pair: longitudinal polarization~(LP) and two transverse polarization~(TP) states.
The LP state is a pure {\it CP}-even eigenstate with helicity amplitude $H_0$, while the TP states are mixtures of {\it CP}-even and {\it CP}-odd states with helicity amplitudes $H_+$ and $H_-$.
The fraction of LP, defined as $f_L = |H_0|^2/(|H_0|^2 + |H_+|^2 + |H_-|^2)$, is measured from the helicity angle distributions of $\rho^\pm$ mesons. The distribution is given by
\begin{eqnarray}
\frac{1}{\Gamma} \frac{d^2\Gamma}{d\cos\theta_{\rho^+} d\cos\theta_{\rho^-}}  =&\frac{9}{4}\Big[\frac{1}{4}(1-f_L)\sin^2\theta_{\rho^+}\sin^2\theta_{\rho^-} \notag \\
&+ f_L \cos^2\theta_{\rho^+} \cos^2\theta_{\rho^-}\Big],
\label{eq:tdcpv}
\end{eqnarray}
where $\theta_{\rho^\pm}$ is the angle in the rest frame of the $\rho^\pm$ meson between the momentum of the $\pi^0$ from the $\rho^\pm \to \pi^\pm \pi^0$ decay and the $B$ flight direction~\cite{ParticleDataGroup:2024xxx}.

Previously, the Belle and \babar experiments measured the branching fraction, $f_L$, and {\it CP} violation parameters for $B^0 \to \rho^+\rho^-$. The results are summarized in Table~\ref{tab:prev} and confirm the dominance of LP.
In this paper, we present a measurement of the branching fraction, the fraction of LP, and time-dependent {\it CP} asymmetries in $B^0 \to \rho^+ \rho^-$ using data from Belle~II. The analysis is presented as follows. In Sec.~\ref{sec:data_set_and_detector} we describe the Belle II detector and data set used; in Sec.~\ref{sec:reconstruction} we discuss event reconstruction and selection of signal candidates; in Sections~\ref{sec:fit_strategy} and ~\ref{sec:fit_results} we describe the fitting procedure and present the fit results; in Sec.~\ref{sec:syst} we discuss systematic uncertainties; in Sec.~\ref{sec:phi2} we perform an isospin analysis to determine the angle $\phi_2$; and in Sec.~\ref{sec:conclusions} we conclude.
\section{Data set and Belle II detector}
\label{sec:data_set_and_detector}

This measurement is based on $(365.4\pm 1.7)$~$\rm{fb}^{-1}$~\cite{Belle2Lum:2024} of data containing $(387\pm 6)\times 10^6$ $\Upsilon(4S)$ resonance decays~\footnote{A small number of non-resonant $e^+e^-\to B\bar{B}$ events are also included in the number of $Y(4S)$ decays; these events are indistinguishable from $e^+e^-\to \Upsilon(4S)\to B\bar{B}$.}.
An additional sample of $42.7\pm0.2$~fb$^{-1}$ accumulated at an energy 60~MeV below the $\Upsilon (4S)$ peak (off-resonance data), which is below the $B \Bbar$ threshold, is used to study backgrounds.
These data samples were taken with the Belle~II detector~\cite{Abe:2010gxa} at the SuperKEKB asymmetric-energy $e^+e^-$ collider~\cite{Akai:2018mbz} in 2019--2022.

The Belle II detector~\cite{Abe:2010gxa} has a cylindrical geometry and includes a two-layer silicon-pixel detector~(PXD) surrounded by a four-layer double-sided silicon-strip detector~(SVD)~\cite{Belle-IISVD:2022upf} and a 56-layer central drift chamber~(CDC). These detectors reconstruct tracks of charged particles. Only one sixth of the second layer of the PXD was installed for the data analyzed here. The symmetry axis of these detectors, defined as the $z$ axis, is almost coincident with the direction of the electron beam. Surrounding the CDC, which also provides $dE/dx$ energy-loss measurements, is a time-of-propagation counter~(TOP)~\cite{Kotchetkov:2018qzw} in the central region and an aerogel-based ring-imaging Cherenkov counter~(ARICH) in the forward region. These detectors provide charged-particle identification. Surrounding the TOP and ARICH is an electromagnetic calorimeter~(ECL) based on CsI(Tl) crystals that primarily provides energy and timing measurements for photons and electrons. Outside of the ECL is a superconducting solenoid magnet. The flux return of the magnet is instrumented with resistive-plate chambers and plastic scintillator modules to detect muons, $K^0_L$ mesons, and neutrons. The magnet provides a 1.5~T magnetic field that is oriented parallel to the $z$ axis.

We use Monte Carlo (MC) simulated events to optimize selection criteria, calculate reconstruction efficiencies, and study sources of background.
The $B \Bbar$ samples are generated with \textsc{EvtGen}~\cite{Lange:2001uf}.
Continuum $e^+e^- \to \qqbar~(q=u,d,s,c)$ and $e^+e^-\to\tau^+\tau^-$ events are generated with KKMC~\cite{Jadach:1999vf}. The fragmentation of $\qqbar$ uses \textsc{Pythia8}~\cite{Sjostrand:2014zea}, and $\tau$ decays are simulated by \textsc{Tauola}~\cite{Skwarnicki:1986xj}.
Final state radiation is simulated by \textsc{Photos}~\cite{Barberio:1993qi}.
\textsc{Geant4}~\cite{Agostinelli:2002hh} is used to simulate the detector response to the passage of particles.
Our simulation includes effects of beam-induced backgrounds~\cite{Lewis:2018ayu,Liptak:2021tog}.
Both the data and simulated events are reconstructed using the Belle II software framework {\text basf2} \cite{Kuhr:2018lps,basf2-zenodo}.
\section{Reconstruction and Event selection} 
\label{sec:reconstruction}
Hadronic \BBbar events are selected online using a hardware trigger based on total energy and charged-particle multiplicity, with an efficiency close to 100\%.
We subsequently reconstruct the $B^0\to\rho^+ \rho^-$ decay from $\rho^+ \to \pi^+\pi^0$ and $\pi^0 \to \gamma \gamma$.
To reject misreconstructed tracks and charged particles from beam-induced background, we require that tracks be within the polar-angle acceptance of the CDC ($17^\circ < \theta < 150^\circ$) and have a distance-of-closest-approach to the $e^+e^-$ interaction point (IP) of less than 0.5~cm in the transverse direction and less than 3.0~cm in the longitudinal($z$) direction.
We select $\pi^\pm$ candidates using a charged particle identification (PID) ratio $\mathcal{L}_{\pi}/(\mathcal{L}_{\pi}+\mathcal{L}_{K})$, where $\mathcal{L}_{\pi(K)}$ is the likelihood for a pion (kaon) hypothesis based mainly on information from the TOP and ARICH detectors, and, for low momentum tracks, the CDC.
The pion-identification efficiency is 97\% and the probability of a kaon being misidentified as a pion is 28\%.

Photons are identified as clusters of ECL crystals having energy deposition above a threshold and not matched to tracks.
We require minimum energies of 60~MeV and 90~MeV in the ECL barrel and 
endcap regions, respectively, where the barrel corresponds to the range $[32.2, 128.7]^\circ$ in polar angle, and the forward and backward endcaps correspond to the ranges $[12.4, 31.4]^\circ$ and $[130.7, 155.1]^\circ$, respectively.
These minimum energy requirements suppress combinatorial background from low-energy photons. A tighter energy requirement is applied to the endcap region due to the higher level of beam-induced backgrounds in this region. To further suppress backgrounds, we require that there be at least two ECL crystals in the ECL cluster, and that the cluster time be within 200 ns of the collision time.

\begin{figure}[htb]
  \centering
  \includegraphics[width=0.48\textwidth]{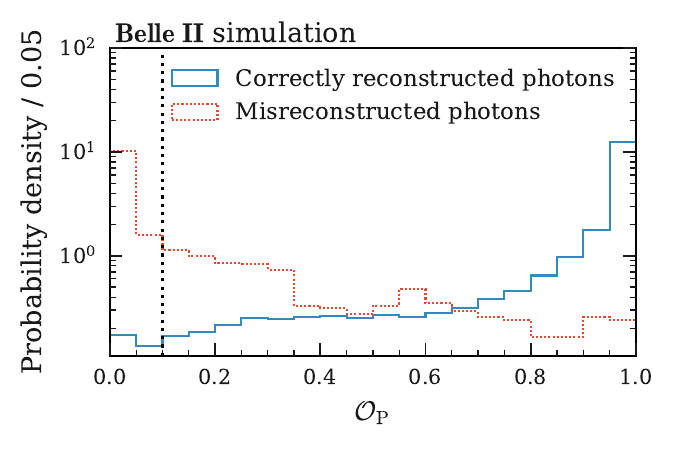}
  \caption{Distribution of the FBDT classifier to distinguish correctly reconstructed photons~(blue solid histogram) from hadronic clusters or splitoffs from charged-particle tracks~(red dotted histogram) in simulation. The black dotted line corresponds to the threshold.}
  \label{fig:pmva}
\end{figure}

To distinguish photons from hadronic clusters or energy deposits split off from hadronic clusters, we train a fast boosted decision-tree (FBDT)~\cite{Belle-II:FBDT} 
with the polar and azimuthal angles, energy, transverse momentum, and ten variables related to the cluster shape and its uncertainty.
The classifier output used to distinguish correctly reconstructed photons is shown in Fig.~\ref{fig:pmva}.
We choose a loose threshold on the FBDT output of $\mathcal{O}_P > 0.1$, which removes 58\% of mis-reconstructed photons while retaining 98\% of signal photons.

We reconstruct $\pi^0$ candidates from pairs of photon candidates with invariant mass in the range $120~\mathrm{MeV/}c^2<m_{\gamma\gamma}<150~\mathrm{MeV/}c^2$. The average mass resolution is approximately 6~MeV/$c^2$.
Additionally, $\pi^0$ candidates must have momenta greater than 210~MeV/$c$. We also require that the difference between the polar angles of the photon momenta be less than 1.3 radians, and the difference between the azimuthal angles be less than 2.5 radians. The $\pi^0$ candidates satisfying these criteria are refit, subject to a $\pi^0$ mass constraint; i.e., the daughter photon momenta are slightly adjusted such that the invariant mass $M_{\gamma \gamma}$ equals $m_{\pi^0}$.

The selected $\pi^\pm$ and $\pi^0$ candidates are combined to form $\rho^\pm$ meson candidates, and we require $0.6~\mathrm{GeV/}c^2<m_{\pi^\pm\pi^0}<1.1~\mathrm{GeV/}c^2$.
A $B^0$ candidate is reconstructed from pairs of $\rho^+ \rho^-$ candidates.
We fit the $B^0$ decay vertex with the TreeFit package~\cite{Krohn:2019dlq}, constraining the $B^0$ to originate from the IP.  
The resulting momenta 
of the decay products adjusted by the fit are used for further analysis.
To further identify $B^0$ decays, we use two kinematic variables: the beam-energy-constrained mass $M_{\mathrm{bc}}\equiv \sqrt{{E_{\mathrm{beam}}^{*}}^2/c^4-{p_{B}^{*}}^2/c^2}$, and the energy difference $\Delta E \equiv E_{B}^{*} - E_{\mathrm{beam}}^{*}$, where $E_{\mathrm{beam}}^{*}$ is the beam energy and $E_{B}^{*}$ ($p_{B}^{*}$) is the energy (momentum) of the $B$ meson, all evaluated in the center-of-mass (c.m.) frame.
Candidate $B$ mesons are required to satisfy $M_{\mathrm{bc}}>5.275$~GeV/$c^2$ and $|\Delta E|<0.15$~GeV.

We determine the decay vertex of $B_{\rm tag}$ using the remaining tracks in the rest of the event~(ROE).
The tracks are required to have at least one hit in each of the PXD, SVD, and CDC detectors and have momenta greater than 50~MeV/$c$. 
Each track must originate from the IP and have impact parameters within 0.5~cm in the transverse direction and within 2.0~cm in the longitudinal direction.
We determine an initial $B_{\rm{tag}}$ decay vertex from a $\chi^2$ fit to all ROE tracks satisfying the above selection criteria; we then remove from the fit, one by one, tracks that contribute most to the $\chi^2$ until the reduced $\chi^2$ is less than four. 
This minimization reduces the impact of displaced vertices due to intermediate charm mesons. 
We calculate $\Delta t_l$ from the distance, $\Delta l$, between the $B_{\it CP}$ vertex and the the $B_{\rm tag}$ vertex along the $\Upsilon(4S)$ boost direction,
\begin{eqnarray}
\Delta t_l =\frac{\Delta l}{c\beta \gamma \gamma_B},
\label{eq:deltat_l}
\end{eqnarray}
where $\beta \gamma= 0.28$ is the Lorentz boost of the $\Upsilon(4S)$ in the lab frame, and $\gamma_B = 1.002$ is the Lorentz boost of the $B$ in the c.m. frame.
The uncertainty on $\Delta t_l$, defined as $\sigma_{\Delta t}$, is estimated event-by-event by the vertex fitter.
We reject poorly reconstructed events by requiring $|\Delta t_l|<15~\mathrm{ps}$ and $\sigma_{\Delta t} < 2.00~\mathrm{ps}$. The quantity $\Delta t_l$ can differ slightly from the $\Delta t$ of Eq.~\ref{eq:tdcpv}, with the difference depending on the direction of the $B^0$ in the c.m. frame. We correct for this small effect in the same manner as done in Ref.~\cite{Belle-II:2023bps}. The flavor of $B_{\rm tag}$ is identified by a flavor-tagging algorithm based on a graph neural network (GNN) that uses properties of charged particles in the ROE~\cite{belleiicollaboration2024new}.
The GNN calculates the flavor $q$ of $B_{\rm tag}$ and a quality factor $r$. The latter ranges from zero for no discriminating power to one for unambiguous flavor assignment.
We do not impose a selection requirement on $r$; rather, we divide the data into seven bins of $r$ and use this binning in our subsequent fitting procedure. The bin boundaries are 0, 0.10, 0.25, 0.45, 0.60, 0.725, 0.875, and 1.

After the above reconstruction, we apply a selection $\cos \theta_{\rho^\pm} < 0.9$ to suppress combinatorial backgrounds from low-momentum $\pi^0$'s.
The continuum background is suppressed using a TabNet classifier~\cite{arik2020tabnet} that distinguishes the difference in topology between continuum events, which tend to be jet-like, and \BBbar events, which tend to be spherical.
We use the following 21 variables calculated in the c.m. frame as input parameters to the TabNet classifier:
the cosine of the angle between the thrust axis~\cite{Brandt:1964sa} of the daughter particles of the $B$ candidate and $z$-axis;
the cosine of the angle between the thrust axes of the daughter particles of the $B$ candidate and all other particles in the ROE;
the cosine of the angle between the thrust axis of all particles in the event and the $z$-axis;
the thrust value of the particles in the ROE;
the cosine of the angle between the $B$ momentum direction and the the $z$-axis;
the cosine of the angle, in the $B$ rest frame, between the $\rho^\pm$ direction and the boost direction of the $B$ in the lab frame; and
modified Fox-Wolfram moments~\cite{Fox:1978vu, Belle:2003fgr}.
We train the classifier using a signal MC simulation sample with $f_L=0.99$ and a \qqbar background MC sample. The TabNet classifier removes more than 99\% of \qqbar background while retaining 38\% of signal according to the simulation.

After the event selection and background suppression, 
14.6\% events have multiple candidates with, an average multiplicity of 2.3.
In events with multiple candidates, we select the candidate with the smallest sum of three reduced-$\chi^2$ values: two from the fits to photon pairs with their invariant mass constrained to the $\pi^0$ mass~\cite{ParticleDataGroup:2024xxx}, and one from the $B$-vertex fit.
According to MC studies, the above criteria choose the correct signal decay in 45\% of multiple-candidate events for LP decays, and in 58\% of such events for TP decays. These rates are a notable improvement over a random selection, for which the corresponding fractions are 37\% and 46\%.

The final selection criteria for photon energy, classifier output $\mathcal{O}_{P}$, angles between the momenta of the two photons, $m_{\gamma\gamma}$, $\pi^0$ momentum, PID, output of the TabNet classifier, and $M_{\rm{bc}}$ are optimized simultaneously by differential evolution~\cite{DE} to maximize a figure-of-merit (FoM) $S/\sqrt{S+B}$, where $S$ and $B$ are the expected $B^0\to\rho^+\rho^-$ signal and background yields, respectively, in a signal-enhanced region defined as $-0.08 ~\mathrm{GeV}<\Delta E< 0.04~\mathrm{GeV}$ and $0.6~\mathrm{GeV/}c^2<m_{\pi^\pm\pi^0}<0.9~\mathrm{GeV/}c^2$.

After the single candidate selection, the efficiencies for correctly reconstructed LP and TP signals calibrated with control samples as described in Section~\ref{sec:syst} are 4.1\% and 7.8\%.
These efficiencies include the branching fraction of $\pi^0 \to \gamma \gamma$. 
Some signal candidates are incorrectly reconstructed, e.g., they contain decay products of $B_{\rm tag}$.
These are referred to as self-crossfeed (SCF) events.
Their $\Delta t$ distributions depend on the number of correctly reconstructed charged pions.
Thus, we divide SCF events into two categories. The first category (``$\rm{SCF_a}$") consists of decays in which both charged pions are correctly reconstructed, while the second category (``$\rm{SCF_{b}}$") consists of decays in which at least one of the charged pions is mis-reconstructed. For $\rm{SCF_a}$ events, the $\Delta t_l$ distribution is unaffected, whereas for $\rm{SCF_b}$ events, the $\Delta t_l$ distribution is smeared.

We check the consistency between data and simulation by reconstructing three control channels. The first is the decay chain $B^+\to D^0\rho^+$, $D^0\to K^-\pi^+\pi^0$, and $\rho^+\to\pi^+\pi^0$, which is used to check the $\Delta E$, $m_{\pi^\pm\pi^0}$, and $\cos\theta_{\rho^\pm}$ distributions.
The same event selections as those used for the signal are applied to the $\pi^\pm$, $\pi^0$, and $\rho^\pm$ candidates. The second control channel is $B^0 \to D^{*-}\pi^+$, $D^{*-}\to\Dzb\pi^-$, and $\Dzb \to K^+\pi^-$, which is used to check the TabNet classifier output and $\Delta t$ distributions.
The third control channel is $B^0\to\rho^+\rho^-$ candidates in the $M_{\rm bc}$ sideband 5.24~${\rm GeV}/c^2<M_{\rm bc}<5.26~{\rm GeV}/c^2$. This sample is used to check the background distributions.

\section{Fit Strategy and Modeling}
\label{sec:fit_strategy}
To measure $\mathcal{B}$, $f_L$, $S$, and $C$, we perform two unbinned maximum-likelihood fits. The first fit is an extended fit to the six observables $\Delta E$, $m_{\pi^{\pm}\pi^{0}}$, $\cos\theta_{\rho^{\pm}}$, and $\mathcal{T}_C$, and determines the parameters $\mathcal{B}$ and $f_L$. The observable $\mathcal{T}_C$ is a transformed output of the TabNet classifier that is distributed uniformly between zero and one for LP signal events. Continuum background events, in contrast, peak at zero in $\mathcal{T}_C$.
The second fit is to the observables $\Delta t$, $q$, and $r$, and determines $S$ and $C$. For the latter fit, the signal probability is calculated event-by-event using the signal and background yields obtained from the first fit.
The values for $S$ and $C$ are determined by an unbinned maximum likelihood fit to the $\Delta t$ distributions for $q=+1$ and $q=-1$ events, in seven bins of $r$.

The components used in the fits are LP and TP signals, their SCF contributions,
continuum, combinatorial and ``peaking" $B \Bbar$ backgrounds, and $e^+e^-\to\tau^+\tau^-$ ($\tau^+\tau^-$) events.
Here, peaking $B \Bbar$ backgrounds refer to charmless $B$ decays that peak in the $\Delta E$ or $m_{\pi^\pm\pi^0}$ distributions. Other $B \Bbar$ backgrounds are categorized as combinatorial $B \Bbar$ backgrounds.

The likelihood for the signal-extraction fit is expressed as
\begin{eqnarray}
\small
\begin{aligned}
\mathcal{L}=
&\frac{\prod_{j} e^{-N_j}}{n_{{\rm tot}}!} \prod_{i=1}^{n_{{\rm tot}}} 
  \sum_{j}\\
&N_j\mathcal{P}_j(\Delta E,~m_{\pi^+\pi^0},~m_{\pi^-\pi^0},~\cos \theta_{\rho^+},~\cos \theta_{\rho^-},~\mathcal{T}_C),
\end{aligned}
\end{eqnarray}
where $N_j$ is the yield for component $j$,
$n_{{\rm tot}}$ is the total number of events, 
and $\mathcal{P}_j$ is the probability density function (PDF) for component $j$. 
The number of SCF events is proportional to the signal yield with a proportionality factor $k$.
The values for the $\mathcal{B}$ and $f_L$ parameters are calculated from the LP and TP signal yields~($N^{\rm LP(TP)}_{\rm sig}$), using
\begin{eqnarray}
  \mathcal{B}  &=& \cfrac{(N_{\rm sig}^{\rm LP}/\epsilon_{\rm LP} + N_{\rm sig}^{\rm TP}/\epsilon_{\rm TP}) }{2 N_{\Upsilon(4S)}f_{00}},\\
  f_L &=& \frac{N_{\rm sig}^{\rm LP}/\epsilon_{\rm LP}}{N_{\rm sig}^{\rm LP}/\epsilon_{\rm LP} + N_{\rm sig}^{\rm TP}/\epsilon_{\rm TP}},
  \label{eq:bfflyield}
\end{eqnarray}
where $\epsilon_i$ are the signal efficiencies,
and $N_{\Upsilon(4S)}$ is the number of $\Upsilon(4S)$ events produced.
The factor $f_{00}$ is the fraction of $B^0\Bzb$ events in $\Upsilon(4S)$ decays. The value of $f_{00}$ is taken from the Heavy Flavor Averaging Group (HFLAV) and equals $0.4861^{+0.0074}_{-0.0080}$~\cite{banerjee2024averagesbhadronchadrontaulepton}. 

The likelihood for the time-dependent {\it CP}-asymmetry fit is 
\begin{equation}
\begin{aligned}
\mathcal{L}(\Delta t, \sigma_{\Delta t},q,r) =
\prod_{i=1}^{n_{{\rm tot}}} \Bigl[
 \sum_{j} f_{j}\mathcal{P}_{j}(\Delta t,\sigma_{\Delta t},q,r)\Bigr],\\
\end{aligned}
\label{eq:deltaT_fitting_pdf}
\end{equation}
where 
$\mathcal{P}_j(\Delta t)$ is the PDF for component $j$, and $f_j$ is the event-by-event component fraction calculated using the result of the signal-extraction fit. 
The fractions $f_j$ are given by 
\begin{equation}
\label{eq:signal_fraction}
\small
f_j = \frac{N_{j}\mathcal{P}_j(\Delta E, m_{\pi^+\pi^0},m_{\pi^-\pi^0}, \cos\theta_{\rho^+},\cos\theta_{\rho^-} )\mathcal{P}_j(r)}{ \sum_k N_k \mathcal{P}_k(\Delta E, m_{\pi^+\pi^0},m_{\pi^-\pi^0}, \cos\theta_{\rho^+},\cos\theta_{\rho^-} )\mathcal{P}_k(r)},
\end{equation}
where $N_j$ is the number of events for the component $j$,
$\mathcal{P}_j(\Delta E, m_{\pi^+\pi^0},m_{\pi^-\pi^0}, \cos\theta_\rho^+,\cos\theta_\rho^- )$ is the same PDF as in the signal-extraction fit, and we also include $\mathcal{P}_j(r)$, which is the PDF for $r$ as a histogram with seven $r$ bins. The PDFs for $r$ are extracted from simulated samples. The variable $\sigma_{\Delta t}$ is utilized as a conditional variable in the resolution function of the $\Delta t$ PDF.
The various PDFs used in the fits are described below and summarized in Table~\ref{tab:models}.

\subsection{Signal-Extraction Fit}
\subsubsection{Correctly reconstructed Signal}\label{sec:model_long_signal}
The $\Delta E$ distribution is described by two bifurcated Gaussian functions with a common mean.
The $\rho^\pm$ lineshape is modeled by a relativistic Breit-Wigner~(BW) function given by
\begin{eqnarray}
  A(m) &=&\frac{p_{\pi}}{m^2 - m_0^2 + im_0\Gamma(m)} 
   \frac{F(p_{\rho})}{F(p^\prime_{\rho})}
  \frac{F(p_{\pi})}{F(p^\prime_{\pi})},\\
    \Gamma(m) &=& \left(\frac{p_{\pi}}{p^\prime_{\pi}}\right)^3 \left(\frac{m_0}{m}\right)\Gamma_0\left[\frac{F(p_{\pi})}{F(p^\prime_{\pi})}\right]^2,
    \label{eq:breit_wigner}
\end{eqnarray}
where $m_0$ and $\Gamma_0$ are the peak position and width, respectively, of the $\rho^\pm$ meson.
The variable $p_P$ is the momentum of the $P$ particle in the rest frame of its parent particle,
and $p^\prime$ is the momentum when the mass $m_{\pi^+\pi^0}$ equals $m_\rho$.
The term $F(p)$ is the Blatt-Weisskopf form factor $F(p) = 1/\sqrt{1+(Rp)^2}$, as described in Ref.~\cite{Blatt:1952ije}. Here, $R$ is the meson radius parameter, which is set to 3~GeV$^{-1}$~\cite{Lange:2001uf}. The PDF for the fit is obtained by $|A(m)|^2$. The parameters $m_0$ and $\Gamma_0$ are estimated by MC. They have different values from the PDG values~\cite{ParticleDataGroup:2024xxx} to account for the effects of detector resolution.

The $\mathcal{T}_{C}$ distribution is parametrized by a linear function whose parameters are determined from the $B^0\to D^{* -}\pi^+$ sample.
The $\cos\theta_{\rho^\pm}$ distribution is modeled using a one-dimensional histogram template that depends on $\Delta E$, to account for a correlation between $\cos\theta_{\rho^\pm}$ and $\Delta E$.
These templates are determined from MC-simulated samples. The $\Delta E$ and $m_{\pi^\pm\pi^0}$ peak positions and widths are calibrated using the $B^+ \to D^0 (\to K^- \pi^+ \pi^0)\rho^+$ control sample, as these decays, like signal decays, have two $\pi^0$'s in the final state. Since $B^+ \to D^0 \rho^+$ decays are longitudinally polarized, as $B^0\to\rho^+\rho^-$ decays essentially are, their $\cos\theta_{\rho^\pm}$ distribution is used to validate the $\cos\theta_{\rho^\pm}$ distribution of signal decays (i.e., the $B^+\to D^0\rho^+$ distributions are compared between data and MC simulation to check for good agreement). The signal yield is floated in the fit.

\subsubsection{Signal Self-crossfeed}
For self-cross-feed signal, the $\Delta E$ distributions are described by a bifurcated Gaussian function.
The lineshape of $m_{\pi^\pm\pi^0}$ is described by the sum of a linear function and a relativistic BW function.
The $\mathcal{T}_{C}$ distribution is modeled by a linear function.
A two-dimensional histogram template is used for $\cos\theta_{\rho^+}$ and $\cos\theta_{\rho^-}$, as these two observables are correlated.
The same calibration factors as those used for correctly reconstructed signal are applied.
The ratios of the SCF yields to the signal yields are fixed to the MC values: $k^{\rm{LP}}_{\mathrm{SCF_a}}=0.19$, $k^{\rm{LP}}_{\mathrm{SCF_b}}=0.16$ and $k^{\rm{TP}}_{\mathrm{SCF}}=0.08$.
The parameters for SCF modeling are determined from MC-simulated samples.
\subsubsection{Continuum}
For continuum events, the $\Delta E$ distribution is described by a quadratic function.
Since continuum events also include $\rho$ resonances, the $m_{\pi^\pm\pi^0}$ distributions are modeled by a sum of a relativistic BW function and a linear function.
As there are correlations in the $\Delta E$-$\mathcal{T}_C$ and $m_{\pi^\pm\pi^0}$-$\cos\theta_{\rho^\pm}$ distributions, one-dimensional histogram templates depending on $\Delta E$ are adopted for modeling the $\mathcal{T}_C$ distribution, and the $m_{\pi^\pm\pi^0}$-$\cos\theta_{\rho^\pm}$ distributions are modeled in a similar way.
These parameters are determined from the MC simulation samples.
The same calibration parameters as those for signals are used for the $m_{\pi^\pm\pi^0}$ peak position and width.
The continuum yield is free to vary.

A mis-modeling of the $\cos\theta_{\rho^\pm}$ distribution for \qqbar MC samples is found in the $M_{\rm{bc}}$ sideband data.
To improve the PDF modeling, we correct the $\pi^\pm$ momentum distribution in the lab frame for the \qqbar MC samples using sideband data, assuming that the $B \Bbar$ background has the same distributions as the simulated sample. The correction factor for the \qqbar distribution is consistent with what is obtained from studying off-resonance. The corrected $\cos\theta_{\rho^\pm}$ PDF is validated using sideband data and found to be consistent.

\subsubsection{Combinatorial $B \Bbar$ backgrounds}
The $\Delta E$ and $m_{\pi^\pm\pi^0}$ distributions are both described by quadratic functions. The $\Delta E$-$\mathcal{T}_C$ and $m_{\pi^\pm\pi^0}$-$\cos\theta_{\rho^\pm}$ distributions are modeled in the same way as continuum background.
These parameters are obtained from MC simulated samples except for the $\Delta E$ shape.
The $\Delta E$ shape parameters and the yield for combinatorial $B \Bbar$ background are floated in the fit.

\subsubsection{Peaking $B \Bbar$ backgrounds}
The decay of $B$ mesons to all-pion final states or to final states with pions and a kaon 
could peak in the fit observables.
The peaking backgrounds are modeled individually, as summarized in Table~\ref{tab:models}. The branching fractions used are measured values when possible and are listed in Table~\ref{tab:peakings:B}.

Most peaking backgrounds that are not yet measured are added together and treated as ``rare peaking" backgrounds.
Exceptions to this are modes with the same final state as $B^0 \to \rho^+ \rho^-$, modes that are expected to have large branching fractions, and modes whose branching fractions can be estimated from similar modes, as described below.
The modeling of these peaking backgrounds is summarized in Table~\ref{tab:models}. The yield of the rare peaking background is floated in the signal extraction fit, while the relative contribution of an individual mode is fixed based on MC simulation.

The yields for the peaking backgrounds that have the same final state as $B^0 \to \rho^+ \rho^-$, such as the decay chain $B^0 \to \rho^{\pm} \pi^{\mp} \pi^0$, $\rho^\pm\to\pi^\pm\pi^0$, $B^0 \to \pi^+ \pi^- \pi^0 \pi^0$ and $B^0 \to a_1^0 \pi^0$, $a_1^0\to \pi^+\pi^-\pi^0$, are floated in the signal-extraction fit. 
The effect of interference is not considered in the nominal fit but is included as a systematic uncertainty.
The $B^0 \to \pi^+ \pi^- \pi^0$ decay contributes with the addition of a $\pi$ from $B_{tag}$, and the PDFs are different from those of the combinatorial $B\Bbar$ background and the signal. The yield of this component is floated in the fit.

The yields for $B^+\to a_1^0\rho^+$ decays are fixed and also summarized in Table~\ref{tab:peakings:B}. The branching fraction of this decay is unmeasured, so we assume it to be 2.5 times the measured branching fraction of $B^\pm \to a^0_1 \pi^\pm$ and assign a 100\% uncertainty.
For $B^0 \to a_1^\pm \rho^\mp$, since the 90\% C.L. upper limit on the branching fraction is $6.1 \times 10^{-5}$, we take this branching fraction to be $(3 \pm 3) \times 10^{-5}$.
The yields for $B \to a_1 \rho$ decays are fixed and also summarized in Table~\ref{tab:peakings:B}.
For the modes peaking in $\Delta E$ or $m_{\pi^\pm\pi^0}$, these PDFs are calibrated as described in Section~\ref{sec:model_long_signal}.

\begin{table*}[htbp]
\renewcommand{\arraystretch}{1.2}
  \caption{ Summary of fit models for each event type. BG: bifurcated-Gaussian, DBG: sum of two bifurcated-Gaussians sharing a common mean, DG: sum of two Gaussians sharing a common mean, BW: Breit-Wigner distribution, P$_{i}$: $i$ th order polynomial function, and exp: an exponential function. The notation $|_{\rm{X}}$ indicates that the correlation with X is considered. The superscript {\it CP} indicates the inclusion of a {\it CP} asymmetry term, and the subscript $\tau_{B^0}$ or $\tau_{\rm{eff}}$ denote the $B^0$ lifetime or an effective lifetime determined from MC simulation. The factor $\mathcal{R}$ represents a resolution function. The yields for the processes in the upper half of the Table are floated while the yields for processes in the lower half are fixed.}
  \label{tab:models}
  \centering 
\begin{tabular*}{\linewidth}{@{\extracolsep{\fill}}llllll}
    \hline \hline
    Process         & $\Delta E$ & $m_{\pi^\pm\pi^0}$ & $\mathcal{T}_{C}$ &$\cos\theta_{\rho^\pm}$ & $\Delta t$\\
    \hline
    LP signal      &DBG& BW   & $\rm{P_{1}}$& Template$|_{\Delta E}$ & ${\exp^{\it CP}_{\tau_{B^0}}}\otimes \mathcal{R}$\\
    LP $\rm{SCF_a}$     &BG & BW + $\rm{P_{1}}$& $\rm{P_{1}}$& Template$|_{\cos\theta_{\rho^\pm}}$ & ${\exp^{\it CP}_{\tau_{B^0}}}\otimes \mathcal{R}$\\
    LP $\rm{SCF_b}$      &BG & BW + $\rm{P_{1}}$& $\rm{P_{1}}$& Template$|_{\cos\theta_{\rho^\pm}}$ & ${\exp^{ {\it CP} }_{\tau_{\rm eff}}}\otimes \mathcal{R}$\\
    TP signal      &DBG& BW   & $\rm{P_{1}}$& Template$|_{\Delta E}$ & ${\exp_{\tau_{B^0}}}\otimes\mathcal{R}$\\
    TP SCF        &BG & BW + $\rm{P_{1}}$& $\rm{P_{1}}$& Template$|_{\cos\theta_{\rho^\pm}}$ & ${\exp_{\tau_{\rm eff}}}\otimes \mathcal{R}$\\
    $\qqbar$      &$\rm{P_{2}}$~&BW + $\rm{P_{1}}$~&Template$|_{\Delta E}$~&Template$|_{m_{\pi^{\pm}\pi^{0}}}$~& DG\\
    $B \Bbar$      &$\rm{P_{2}}$~&$\rm{P_{2}}$~  &Template$|_{\Delta E}$~&Template$|_{m_{\pi^{\pm}\pi^{0}}}$~& ${\exp_{\tau_{\rm eff}}}\otimes \mathcal{R}$\\
    $B^0\to\rho^{\pm}\pi^{\mp}\pi^0$     & DBG & Template$|_{m_{\pi^{\pm} \pi^{0}}}$ & Template$|_{\Delta E}$~&Template$|_{\cos\theta_{\rho^\pm}}$ & ${\exp_{\tau_{\rm eff}}}\otimes \mathcal{R}$\\
    $B^0\to\pi^+\pi^-\pi^0\pi^0$ & DBG & $\rm{P_{1}}$& Template$|_{\Delta E}$ & Template$|_{\Delta E}$ & ${\exp_{\tau_{\rm eff}}}\otimes \mathcal{R}$\\
    $B^0\to a_{1}^0 \pi^0$       & DBG & Template$|_{m_{\pi^{\pm} \pi^{0}}}$ & Template$|_{\Delta E}$ & Template$|_{\cos\theta_{\rho^\pm}}$ & ${\exp_{\tau_{\rm eff}}}\otimes \mathcal{R}$\\
    $B^0\to\pi^{+}\pi^{-}\pi^{0}$&$\rm{P_{2}}$~&BW + $\rm{P_{1}}$~&Template$|_{\Delta E}$~&Template~& ${\exp_{\tau_{\rm eff}}}\otimes \mathcal{R}$\\
    rare peaking backgrounds     & DBG & BW + $\rm{P_{1}}$& $\rm{P_{1}}$& Template$|_{\cos\theta_{\rho^\pm}}$ & ${\exp_{\tau_{\rm eff}}}\otimes \mathcal{R}$\\
    \hline 
    $\tau^+\tau^-$&$\rm{P_{2}}$~&BW + $\rm{P_{1}}$~&exp + $\rm{P_{1}}$&Template~&DG\\
    $B^0\to\rho^{\pm}\pi^{\mp}$  & $\rm{P_{2}}$ & BW + $\rm{P_{1}}$& $\rm{P_{1}}$& Template$|_{m_{\pi^{\pm} \pi^{0}}}$ & ${\exp_{\tau_{\rm eff}}}\otimes \mathcal{R}$\\
    $B^0\to a_{1}^\pm \pi^\mp$       & DBG & Template$|_{m_{\pi^{\pm} \pi^{0}}}$ & $\rm{P_{1}}$& Template & ${\exp_{\tau_{\rm eff}}}\otimes \mathcal{R}$\\
    $B^0\to a_1^{\pm}\rho^{\mp}$ &exp + $\rm{P_{1}}$& BW + $\rm{P_{1}}$& Template$|_{\Delta E}$ & Template & ${\exp_{\tau_{\rm eff}}}\otimes \mathcal{R}$\\
    $B^0\to K^{*+} \rho^- $       & DBG & Template$|_{m_{\pi^{\pm} \pi^{0}}}$ & $\rm{P_{1}}$& Template & ${\exp_{\tau_{\rm eff}}}\otimes \mathcal{R}$\\
    $B^0\to K_{0}^*(1430)^+\rho^- $   & $\rm{P_{2}}$ & Template$|_{m_{\pi^{\pm} \pi^{0}}}$ & $\rm{P_{1}}$& Template & ${\exp_{\tau_{\rm eff}}}\otimes \mathcal{R}$\\
    $B^+ \to\rho^+\pi^0$    & $\rm{P_{2}}$ & Template$|_{m_{\pi^{\pm} \pi^{0}}}$ & $\rm{P_{1}}$& Template$|_{m_{\pi^{\pm} \pi^{0}}}$ & ${\exp_{\tau_{\rm eff}}}\otimes \mathcal{R}$\\
    $B^+ \to\rho^+\rho^0$   & DBG & BW + $\rm{P_{1}}$& $\rm{P_{1}}$& Template$|_{m_{\pi^{\pm} \pi^{0}}}$ & ${\exp_{\tau_{\rm eff}}}\otimes \mathcal{R}$\\
    $B^+ \to a_1^{0}\pi^+$  & DBG & BW + $\rm{P_{1}}$& $\rm{P_{1}}$& Template$|_{m_{\pi^{\pm} \pi^{0}}}$ & ${\exp_{\tau_{\rm eff}}}\otimes \mathcal{R}$\\
    $B^+ \to a_1^+\pi^0$    & DBG & BW + $\rm{P_{1}}$& $\rm{P_{1}}$& Template & ${\exp_{\tau_{\rm eff}}}\otimes \mathcal{R}$\\
    $B^+ \to a_1^0 \rho^+$   &exp + $\rm{P_{1}}$& BW + $\rm{P_{1}}$& Template$|_{\Delta E}$ & Template & ${\exp_{\tau_{\rm eff}}}\otimes \mathcal{R}$\\
    $B^+ \to K^+ \pi^- \pi^+$         & $\rm{P_{2}}$ & BW + $\rm{P_{1}}$& $\rm{P_{1}}$& Template & ${\exp_{\tau_{\rm eff}}}\otimes \mathcal{R}$\\
    $B^+ \to K_{0}^*(1430)^+ \pi^0$& BG & $\rm{P_{1}}$& $\rm{P_{1}}$& Template$|_{m_{\pi^{\pm} \pi^{0}}}$ & ${\exp_{\tau_{\rm eff}}}\otimes\mathcal{R}$\\
    
    \hline \hline
  \end{tabular*}
\end{table*}
\begin{table}[h]
  \renewcommand{\arraystretch}{1.2}
  \centering
  \caption{
  Peaking backgrounds from $B^0$ and $B^+$ decays, their branching fractions, and the number of events expected in 365.4 $\rm {fb^{-1}}$ of data. We use PDG values~\cite{ParticleDataGroup:2024xxx} for modes that have been measured. For modes that are unmeasured, we assign a 100\% uncertainty to the estimated branching fractions.
  } 
\begin{tabular*}{\linewidth}{@{\extracolsep{\fill}}lcr}
      \hline \hline
      Decay mode \hspace{50pt} & $\mathcal{B} \, \, [10^{-6}]$& $N_{\rm{bg}}$\\
      \hline 
      $B^0\to \rho^\pm \pi^\mp$ & $23.0\pm2.3$ & 5.0 \\
      $B^0\to a_{1}^\pm \pi^\mp$ & $26\pm5$ & 8.1 \\
      $B^0\to a_{1}^\pm \rho^\mp$ & $30\pm30$ & 8.7 \\
      $B^0\to K^{*+} \rho^-$ & $10.3\pm2.6$ & 9.9\\
      $B^0\to K_0^*(1430)^+ \rho^-$ & $28\pm12$ & 10.2\\
      \hline 
      $B^+\to \rho^+ \pi^0$ & $10.9\pm1.4$ & 39.0 \\
      $B^+\to \rho^+ \rho^0$ & $24.0\pm1.9$ & 21.1 \\
      $B^+\to a_{1}^0 \pi^+$ & $20\pm6$ & 5.5 \\
      $B^+\to a_{1}^+ \pi^0$ & $26\pm7$ & 16.5 \\
      $B^+\to a_{1}^0 \rho^+$ & $50\pm50$ & 18.2 \\
      $B^+\to K^+ \pi^- \pi^+$ & $51\pm29$ & 1.2 \\
      $B^+\to K_0^*(1430)^+\pi^0$ & $11.9_{-2.3}^{+2.0}$ & 4.3 \\
      \hline \hline
    \end{tabular*}
  \label{tab:peakings:B}
\end{table}

\subsubsection{$\tau^+\tau^-$ background}
The $\tau^+\tau^-$ background is suppressed by the selection with the TabNet classifier.
The remaining events arise mostly from combinations of three decays: $\tau^-\to\pi^-\pi^+\pi^-\pi^0\nu_\tau$, $\tau^-\to \pi^-\pi^0\nu_\tau$, and $\tau^-\to\pi^-\pi^0\pi^0\nu_\tau$, which account for more than 90\% of the $\tau^+\tau^-$ background. 
The $\Delta E$ distribution for $\tau^+\tau^-$ background is modeled by a quadratic function, and $m_{\pi^\pm\pi^0}$ is modeled by the sum of a relativistic BW function and a linear function.
The $\mathcal{T}_{C}$ distribution is modeled by an exponential function.
The $\cos\theta_{\rho_\pm}$ distributions are described by one-dimensional histogram templates. These parameters are obtained from the MC simulation samples.
In the fit, the yield is fixed to the expectation from MC simulation, $40.2$~events.

\subsection{Time-dependent CP-asymmetry fit}
\subsubsection{Correctly reconstructed Signal}
Similar PDFs are used for the LP and TP signal events.
Since the TP signal decay includes contributions from both {\it CP}-even and {\it CP}-odd states, our baseline fit assumes that  {\it CP}-violating effects cancel out in the TP components; thus, in the PDF function for $\Delta t$, both $S$ and $C$ are set to zero. Possible nonzero values for the TP component are considered when evaluating systematic uncertainties.

The PDF for the $\Delta t$ distribution of LP $B^0\to\rho^+\rho^-$ decays is
\begin{equation}
  \begin{aligned}
    \mathcal{P}(\Delta t,\bar{t},q)=&\frac{1}{4\tau_{B^0}}
    \exp{\left(\frac{-2\bar{t}}{\ \ \tau_{B^0}}\right)}
    \Bigl\{  1-q\Delta w_r + \\&qa^{\rm{tag}}_{\varepsilon,r}( 1 - 2 w_r )\\&
    +[q(1-2 w_r)+a^{\rm{tag}}_{\varepsilon,r}(1-q \Delta w_r)]\\&[S \sin (\Delta m_d \Delta t) - C \cos (\Delta m_d \Delta t)] \Bigr\},
  \end{aligned}
\end{equation}
where 
$\bar{t}$ is the average of the $B_{\rm{sig}}$ and $B_{\rm{tag}}$ decay times,
$w_r$ is the wrong flavor-tag probability for bin $r$,
$\Delta w_r$ is the difference in wrong tag probabilities between $B^0_{\rm{tag}}$ and $\Bzb_{\rm{tag}}$ for bin $r$, 
and $a^{\rm{tag}}_{\varepsilon,r}$ is the asymmetry in $B^0$ and $\Bbar^0$ flavor-tagging efficiencies for bin $r$~\cite{belleiicollaboration2024new}. 
We integrate out the $\bar{t}$ dependence, which is related to the angular distribution of $B_{\rm sig}$ in the $\Upsilon(4S)$ rest frame~\cite{Belle-II:2023bps}.

The vertex resolution broadens the $\Delta t$ distribution relative to the true distribution. To account for this broadening, we use the same resolution function as in Ref.~\cite{Belle-II:2023bps} except for the $f_{\rm tail}$ modeling. The $f_{\rm tail}$ model is replaced with the sum of a constant and a bifurcated Gaussian having $\sigma_{\Delta t}$ as a parameter.
The resolution function parameters are calibrated by fitting a $B^0\to D^{*-}\pi^+$ control sample with flavor tagging parameters fixed to values from Ref.~\cite{belleiicollaboration2024new}.

\subsubsection{Signal Self-crossfeed}
The correct {\it CP} violation parameters can be extracted from LP $\rm {SCF_a}$ events, as the $B$ decay position (determined from the trajectories of the two charged pions) is correctly reconstructed.
Thus, the $\Delta t$ PDF for correctly reconstructed signal is used for the LP $\rm{SCF_a}$ events with shared {\it CP} violation parameters.
In contrast, the $\Delta t$ distribution for the LP $\rm{SCF_b}$ events is biased due to the contamination of charged tracks from $B_{\rm tag}$.
The $\Delta t$ PDF is modeled using the PDF for correctly reconstructed LP signal with an effective lifetime determined from MC-simulated samples.

For the TP SCF events, $\rm{SCF_a}$ and $\rm{SCF_b}$ are modeled with the same $\Delta t$ PDF. This PDF is the same as that used for the correctly reconstructed TP signal but with an effective lifetime determined from simulation.

\subsubsection{Continuum} \label{event_model:continuum}
The $\Delta t$ PDF for continuum background is modeled by the sum of two Gaussian functions with mean parameters set to zero.
The standard deviations and relative fractions of the Gaussians are determined from off-resonance data.

\subsubsection{Combinatrial $B \Bbar$ backgrounds}
The $\Delta t$ PDF for combinatorial $B \Bbar$ backgrounds is modeled by an exponential function with an effective lifetime convolved with the resolution function. The effective lifetime is determined to be $1.39\pm0.02$~ps from the simulation.
The effective lifetime for the combinatorial $B \Bbar$ backgrounds is determined from the sideband data, which is consistent with the MC result.

\subsubsection{Peaking $B \Bbar$ backgrounds}
The $\Delta t$ PDFs for each peaking background and rare peaking background are modeled with the same functional form as the combinatorial $B \Bbar$ backgrounds but with a mode-specific effective lifetime as determined from MC-simulated samples.
The {\it CP} violation parameters are fixed to world average values if measurements are available; or fixed to zero otherwise.
The {\it CP} violation parameters used for the fit are summarized in Table~\ref{tab:peaking_cpv}.
\begin{table}[htbp]
    \centering
    \caption{{\it{CP}} violation parameters of the peaking backgrounds used for the fit. Here, $A_{\it CP}$ is the direct {\it CP} violation parameter for flavor specific modes. We use PDG values for the measured modes~\cite{ParticleDataGroup:2024xxx} while we assign $\pm50$\% uncertainties for the decays that are not measured yet. 
    }
    \begin{tabular}{lccc}
    \hline \hline
    Decay mode & $S$ & $C$&$A_{\it CP}$\\
    \hline
$B^0\to\rho^\pm\pi^\mp\pi^0$  &  0.0 $\pm$ 0.5  &  0.0 $\pm$ 0.5 &--- \\
$B^0\to\pi^+\pi^-\pi^0\pi^0$  &  0.0 $\pm$ 0.5  &  0.0 $\pm$ 0.5 &--- \\
$B^0\to a_{1}^0 \pi^0$  &  0.0 $\pm$ 0.5  &  0.0 $\pm$ 0.5 &--- \\
$B^0\to\pi^{+}\pi^{-}\pi^{0}$  &  0.0 $\pm$ 0.5  &  0.0 $\pm$ 0.5 &--- \\
$B^0\to K_{S}^0\pi^0\pi^0$  &  0.89 $\pm$ 0.30  &  $-$0.21 $\pm$ 0.20 &--- \\
$B^0\to\rho^{\pm}\pi^{\mp}$  &  0.05 $\pm$ 0.07  &  $-$0.03 $\pm$ 0.07 &--- \\
$B^0\to a_{1}^\pm \pi^\mp$  &  $-$0.2 $\pm$ 0.4  &  $-$0.05 $\pm$ 0.11 &--- \\
$B^0\to a_1^{\pm}\rho^{\mp}$  &  0.0 $\pm$ 0.5  &  0.0 $\pm$ 0.5 &--- \\
$B^0\to  K^{*+} \rho^-$ &  ---  &  ---& 0.21 $\pm$ 0.15 \\
$B^0\to K_{0}^{*}(1430)^{+}\rho^- $  &  ---  &  ---& 0.0 $\pm$ 0.5 \\

\hline
$B^+\to\rho^{+}\pi^0$  &  ---  &  ---& 0.03 $\pm$ 0.10 \\
$B^+\to\rho^{+}\rho^0$  &  ---  &  ---& $-$0.05 $\pm$ 0.05 \\
$B^+\to a_1^{0}\pi^{+}$  &  ---  &  ---& 0.0 $\pm$ 0.5 \\

$B^+\to a_1^{+}\pi^0$  &  ---  &  ---& 0.0 $\pm$ 0.5 \\
$B^+\to a_1^0\rho^{+} $  &  ---  &  ---& 0.0 $\pm$ 0.5 \\
    \hline \hline
    \end{tabular}
    \label{tab:peaking_cpv}
\end{table}

\subsubsection{$\tau^+\tau^-$ background}
The $\Delta t$ PDF is modeled by a sum of two Gaussian functions.
The means, standard deviations, and relative fractions are fixed to values from MC simulation.

\subsubsection{Validation of $\Delta t$ PDF}
To validate the $\Delta t$ resolution function, we use a $B^0\to D^{*-}\pi^+$ control sample to fit for parameters $\tau_{B^0}$ and $\Delta m_d$. We set $S=0$, $C=-1$, and $q = q_{\rm tag} \cdot q_{\rm sig}$, where $q_{\rm{tag}}$ and $q_{\rm{sig}}$ represent the flavors of the tag-side and signal side.
The results are $\tau_{B^0}=1.523 \pm 0.033$~ps and $\Delta m_d=0.507\pm0.017$~ps${}^{-1}$, where the uncertainties are statistical.
We further check the $\Delta t$ resolution function using flavor-untagged $B^0\to\rho^+\rho^-$ candidates and obtain $\tau_{B^0} = 1.41\pm0.13$~ps, which is also consistent with the world average value.
These results are in excellent agreement with the world average values of $\tau_{B^0} = 1.517\pm0.004$~ps and $\Delta m_d = 0.5069 \pm 0.0019\rm{~ps^{-1}}$~\cite{ParticleDataGroup:2024xxx}.
We also perform a time-dependent {\it CP}-asymmetry fit to $B^0\to\rho^+\rho^-$ candidates, randomly assigning flavor $\pm1$ to $q$.
We obtain $S = -0.070\pm0.186$ and $C = -0.079\pm0.121$, which are consistent with zero as expected.

\section{Fit results}
\label{sec:fit_results}
\begin{figure*}[htbp]
  \centering
 \includegraphics[width=1\textwidth]{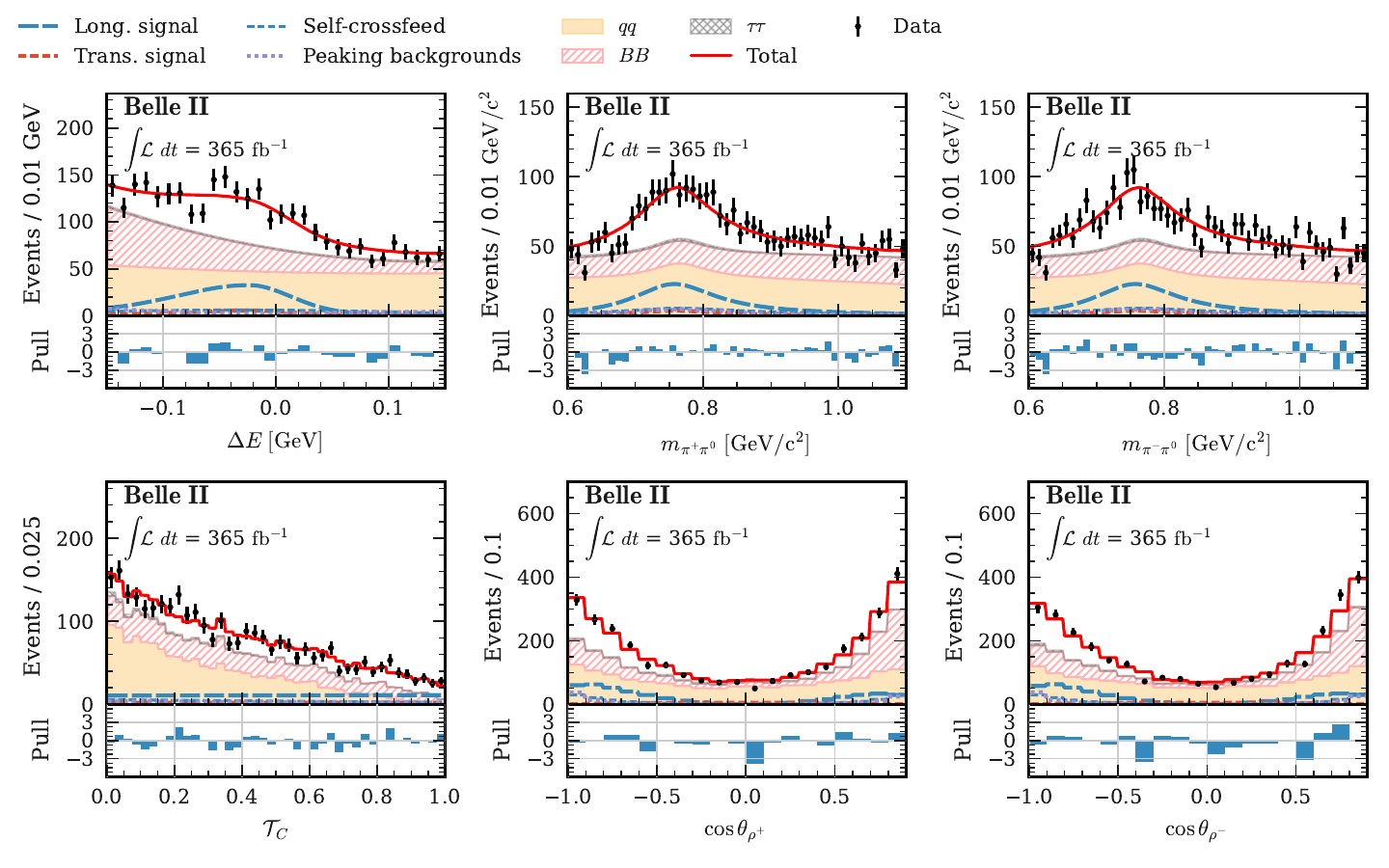}
  \caption{Distributions for $\Delta E$~(top left),  $m_{\pi^\pm\pi^0}$~(top center, top right), $\mathcal{T}_C$~(bottom left), and $\cos\theta_{\rho^\pm}$~(bottom center, bottom right). The points with error bars represent the data, the solid red curves show the sum of all contributions, the long-dashed blue curves show the LP signal, the short-dashed red curves show the TP signal, the short-dashed blue curves show the sum of LP and TP SCF, the dotted purple curves represent peaking backgrounds. The hatched red histograms show the $B \Bbar$ background, the shaded (orange) histograms show the continuum events, and the cross-hatched (black) histograms represent the $\tau^+\tau^-$ background.
  }
  \label{fig:bffit}
\end{figure*}

We first measure the branching fraction and fraction of longitudinal polarization. For this fit, we float $\mathcal{B}$ and $f_L$ as well as the yields of combinatorial $B \Bbar$, \qqbar, \mbox{$B^0\to\rho^{\pm}\pi^{\mp}\pi^0$}, \mbox{$B^0\to\pi^+\pi^- \pi^0\pi^0$}, \mbox{$B^0\to a_1^0\pi^0$}, \mbox{$B^0\to\pi^+\pi^-\pi^0$}, rare peaking backgrounds, and the $\Delta E$ shape for combinatorial $B \Bbar$ backgrounds. 
Figure~\ref{fig:bffit} shows projections of the fit result overlaid on the $\Delta E$, $m_{\pi^\pm\pi^0}$, $\mathcal{T}_C$, and $\cos\theta_{\rho^\pm}$ distributions. The fit result and the data are in good agreement.
We obtain
\begin{align}
  \mathcal{B}(B^0\to\rho^+\rho^-) &= \left(2.89 ^{+0.23}_{-0.22}\right) \times 10^{-5},\\
  f_{L} &= 0.921 ^{+0.024}_{-0.025},
\end{align}
where the uncertainties are statistical only. The statistical correlation is $-0.11$. 
The signal and background yields determined by the fit are summarized in Table~\ref{tab:bffit-result}.

\begin{table}[ht]
\renewcommand{\arraystretch}{1.3}
  \caption{Fit results for signal and background yields. The uncertainties are statistical.}
  \label{tab:bffit-result}
  \centering
  \begin{tabular*}{\linewidth}{@{\extracolsep{\fill}}lr}
    \hline \hline
                 & $N$ \\
    \hline
    LP signal &  ${436.3}^ {+34.2}_{-33.5}$ \\
TP signal &  ${65.4}^ {+24.3}_{-22.6}$ \\
LP signal (SCF) &  ${151.0}^ {+11.9}_{-11.6}$ \\
TP signal (SCF) &  ${5.6}^ {+\phantom{1}1.9}_{-\phantom{1}1.8}$ \\
\qqbar &  ${1410.2}^ {+76.5}_{-75.7}$ \\
Combinatorial $B \Bbar$ &  ${849.2}^ {+73.3}_{-72.3}$ \\
$B^0\to\rho^+\pi^-\pi^0$ &  ${44.9}^ {+93.3}_{-91.9}$ \\
$B^0\to\pi^+\pi^-\pi^0\pi^0$ &  ${-98.0}^ {+62.2}_{-60.2}$ \\
$B^0\to\pi^{+}\pi^{-}\pi^{0}$ &  ${-1.2}^ {+18.8}_{-15.9}$ \\
$B^0\to a_{1}^0 \pi^0$ &  ${32.0}^ {+28.6}_{-26.5}$ \\
Rare peaking backgrounds  &  ${-31.9}^ {+45.3}_{-42.9}$ \\
    \hline \hline
  \end{tabular*}
\end{table}

\begin{figure*}
    \centering
    \includegraphics[width=\linewidth]{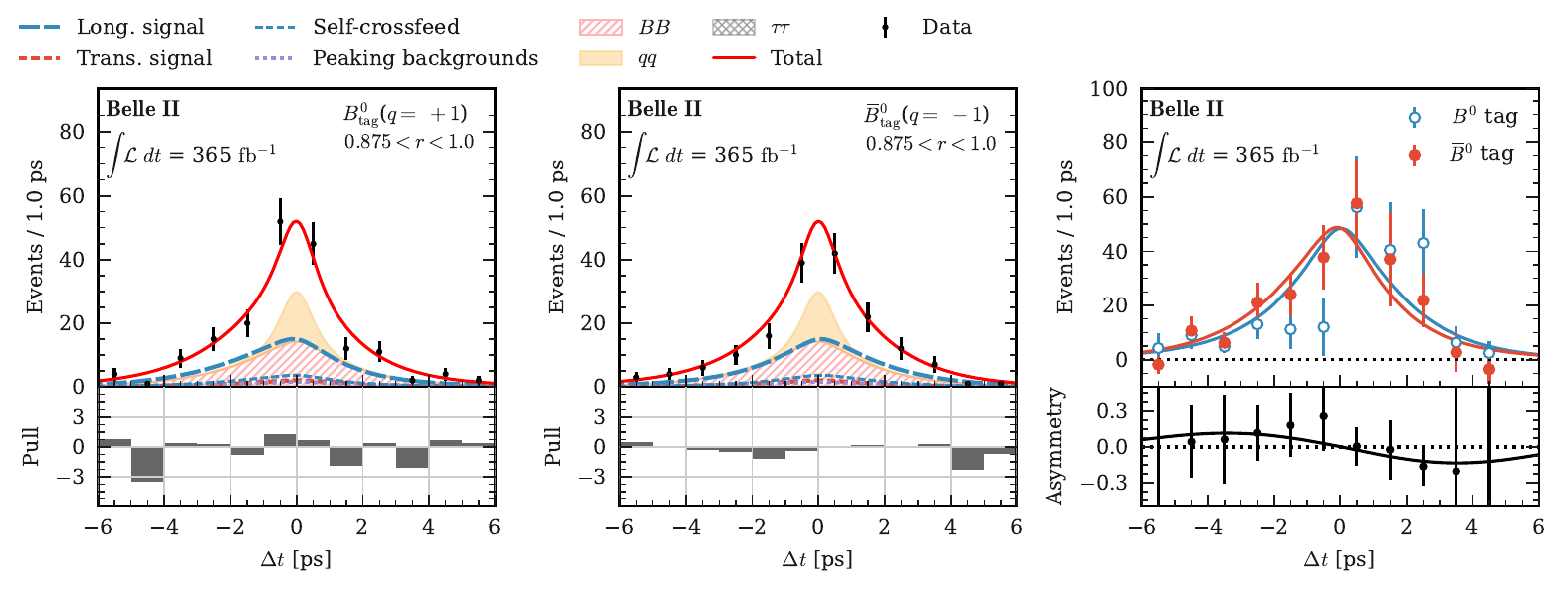}
\caption{Distributions for $\Delta t$ of $B^0_{\text{tag}}$ in $0.875 < r < 1.0$ (left), $\Delta t$ of $\Bzb_{\text{tag}}$ in $0.875 < r < 1.0$  (center), and background-subtracted asymmetry using the ${}_{s}\mathcal{P}lot$ technique~\cite{Pivk:2004ty}. The points with error bars represent the data and the curves show the fit result. The sWeights are calculated using $\Delta E$, $m_{\pi^\pm\pi^0}$, $cos\theta_{\rho^\pm}$, and $qr$.
}
    \label{fig:dtfit}
\end{figure*}

We subsequently perform a second fit to extract the {\it CP} violation parameters.
Figure \ref{fig:dtfit} shows the $\Delta t$ distributions with the fit result superimposed.
The results are
\begin{align}
  S &= -0.26\pm0.19\\
  C &= -0.02\pm0.12,
\end{align}
where the uncertainties are statistical only and include
uncertainties due to statistical uncertainties in $\cal{B}$ and $f_L$.
These are obtained by varying $\cal{B}$ and $f_L$ by their uncertainties, accounting for their correlation. The statistical correlation is $-0.06$.

\section{Systematic uncertainties}
\label{sec:syst}
We consider systematic uncertainties in $\mathcal{B}, f_{L}, S, $ and $C$, which are listed in Tables~\ref{tab:total_uncertaintyBF} and Table~\ref{tab:total_uncertaintyCPV}. The major sources of systematic uncertainties are the $\pi^0$ efficiency for $\mathcal{B}$, fit bias and data-MC mis-modeling for $f_L$, possible $CP$ asymmetries in backgrounds, and the resolution function for $S$ and $C$.

\subsection{Signal-Extraction Fit}
\label{sec:syst_signal_extraction}
Table~\ref{tab:total_uncertaintyBF} summarizes the statistical and systematic uncertainties of the signal-extraction fit.
The signal efficiency is determined based on the MC-simulated samples, corrected using the data-MC ratio of the tracking, $\pi^0$, PID, and TabNet classifier efficiencies. These ratios are evaluated using control samples.

The uncertainty of charged track finding is evaluated using $e^+e^- \to \tau^+\tau^-$ data, in which one $\tau$ decays as $\tau^+\to \ell^+\neul\neutb$ and the other decays as $\tau^+\to\pi^-\pi^+\pi^-\neut$. The data-MC ratios of the tracking efficiency are $0.9999\pm0.0029$ for the LP signal and $0.9999\pm0.0027$ for the TP signal. 
The neutral-pion efficiency is studied using $D^0\to K^- \pi^+\pi^0$, $D^0\to K^-\pi^+$, and $B^-\to D^{*0} \pi^- $ followed by $D^{*0}\to D^0 \pi^0 $ and $D^0\to K^-\pi^+$. The data-MC ratios of the efficiency are {$1.011\pm0.039$} for the LP signal and $1.040\pm0.038$ for the TP signal. 
We obtain a systematic uncertainty associated with the charged-pion identification efficiency using pions from the decay chain $D^{*+}\to D^0 \pi^+$ followed by $D^0\to K^-\pi^+$. The data-MC ratios of the charged-pion identification efficiency are $0.9946\pm0.0004$ for the LP signal and $0.9934\pm0.0004$ for the TP signal. 
The efficiency of the TabNet classifier and the shape of $\mathcal{T}_{C}$ are evaluated using a $B^0\to D^{*-}\pi^+$ control sample. The data-MC ratio of the efficiency is $1.082\pm0.031$.
The systematic uncertainties associated with the efficiency in $\mathcal{B}$ and $f_{L}$ are estimated by varying the efficiency by $\pm1\sigma$.
The statistical uncertainty of the efficiency due to the MC sample size is included in the systematic uncertainty.

The uncertainty due to the single candidate selection is estimated by comparing the results with a random candidate selection and with the nominal one. 
The fractions of SCF events are fixed to the MC expectation and we treat the deviation from the nominal value as the uncertainty, with the mean of $\mathcal{B}$ and $f_L$ obtained from MC ensemble tests varying SCF fraction to the signal by $\pm$ 20\%.
The uncertainties due to peaking backgrounds are evaluated by changing their yields by the fractional uncertainties in their branching fractions, as listed in Table~\ref{tab:peakings:B}.
The uncertainties due to the yields of $\tau^+\tau^-$ backgrounds are evaluated by varying these yields by $\pm100\%$ from their nominal values (obtained from MC simulation), as the phase-space distribution of many of these modes is unknown (e.g., the dominant mode $\tau^-\to\pi^-\pi^+\pi^-\pi^0\nu_\tau$).

The uncertainties due to the signal, \qqbar, $B \Bbar$, $\tau^+\tau^-$, and peaking background modeling are estimated by changing the PDF shape parameters by their uncertainties. The peak positions and widths for $\Delta E$ and $m_{\pi^\pm\pi^0}$ are calibrated using a $B^+\to \bar{D}^0\rho^+$ sample. 
The shifts in peak positions and the data-simulation ratios of the width are $-7.6\pm0.46~\rm{MeV}$, $1.141\pm0.015$ for the $\Delta E$ distribution, and $-9.82\pm0.67~\rm{MeV}/c^2$, $1.025\pm0.014$ for the $m_{\pi^\pm\pi^0}$ distribution.
We also include the uncertainty in the PDF shapes due to the limited MC sample size.
The uncertainties due to interference with $B^0\to\pi^+\pi^-\pi^0\pi^0$, $B^0\to\rho^\pm\pi^\mp \pi^0$ , $B^0\to a_1^0\pi^0$, and $B^0\to a_1^\pm\pi^\mp$ decays, which have the same final-state particles as $B^0\to\rho^+\rho^-$,
is evaluated using simulated datasets, changing the strong phases relative to the signal from zero to $2\pi$ radians assuming the branching fractions of the background modes to be $10^{-5}$.
We repeat the fits with different interference samples and take the standard deviation of the results as the uncertainty.

The systematic uncertainty due to the mis-modeling of MC samples is estimated by changing the $\cos\theta_{\rho^\pm}$ PDF for the combinatorial $B \Bbar$ and \qqbar backgrounds by the differences observed between MC and data events in the $M_{\rm bc}$ sideband.
We estimate the effect of mis-modeling in the $\cos\theta_{\rho^\pm}$  PDF for signal and peaking backgrounds by varying the PDFs by their MC-data differences as measured for the $B^\pm \to D^0\rho^\pm$ control sample.

To check for a possible fit bias, we perform an ensemble test using the MC-simulated samples.
The observed bias is included as a systematic uncertainty.
The uncertainty due to $N_{\Upsilon(4S)}$ is included,  as well as the uncertainty in $f_{00}$ obtained in~Ref.\cite{banerjee2024averagesbhadronchadrontaulepton}.

\begin{table}[htbp]
\renewcommand{\arraystretch}{1.2}
  \centering
  \caption{Systematic uncertainties for $\mathcal{B}$ and $f_{L}$. Relative uncertainties are shown for $\mathcal{B}$.}
\begin{tabular*}{\linewidth}{@{\extracolsep{\fill}}lcc}
  \hline \hline
    Source  & $\mathcal{B} \, \, [\%]$ & $f_{L}[10^{-2}]$\\
    \hline
Tracking & $\pm 0.54$ & --- \\
$\pi^0$ efficiency  & $\pm 7.67$ & --- \\
PID & $\pm 0.08$ & --- \\
$\mathcal{T}_C$ & $\pm 2.87$ & --- \\
MC sample size & $\pm 0.24$ &$\pm 0.2$ \\
Single candidate selection & $\pm 0.55$ &$\pm 0.3$ \\
SCF ratio & $^{ +2.97 }_{-2.45}$ &$^{ +0.2 }_{-0.3}$ \\
$\mathcal{B}$'s of peaking backgrounds & $^{ +0.94 }_{-0.98}$ &$\pm 0.1$ \\
$\tau^{+}\tau^{-}$ background yield & $^{ +0.65 }_{-0.69}$ &$\pm 0.0$ \\
Signal model & $^{ +1.14 }_{-2.02}$ &$\pm 0.2$ \\
$q\bar{q}$ model & $^{ +0.49 }_{-0.51}$ &$^{ +0.1 }_{-0.2}$ \\
$B\bar{B}$ model & $^{ +1.00 }_{-0.40}$ &$^{ +0.3 }_{-0.1}$ \\
$\tau^{+}\tau^{-}$ model & $^{ +0.17 }_{-0.26}$ &$^{ +0.0 }_{-0.1}$ \\
Peaking model & $^{ +1.37 }_{-1.01}$ &$^{ +0.3 }_{-0.5}$ \\
Interference & $\pm 1.20$ &$\pm 0.5$ \\
Data-MC mis-modeling & $^{ +3.51 }_{-1.70}$ &$^{ +0.8 }_{-0.3}$ \\
Fit bias & $\pm 1.03$ &$\pm 1.2$ \\
$f_{00}$ & $^{ +1.67 }_{-1.50}$ & --- \\
$N_{\Upsilon(4S)}$ & $\pm 1.45$ & --- \\
\hline \hline
Total systematic uncertainty & $^{ +10.10 }_{-9.51}$ &$^{ +1.7 }_{-1.5}$ \\
Statistical uncertainty   & $^{ +7.95 }_{-7.61}$  &$^{ +2.4 }_{-2.5}$   \\
\hline \hline
  \end{tabular*}
  \label{tab:total_uncertaintyBF}
\end{table}

\subsection{Time-Dependent CP-asymmetry fit}
Table~\ref{tab:total_uncertaintyCPV} summarizes the statistical and systematic uncertainties of the time-dependent {\it CP}-asymmetry fit. 
The systematic uncertainties due to sources that contribute to both the signal extraction and {\it CP} asymmetry fits are obtained by repeating the signal extraction fit as described in Section~\ref{sec:syst_signal_extraction}, and then repeating the {\it CP}~asymmetry fit with the modified signal fraction.
The following uncertainties are included: branching fractions for peaking backgrounds, $\tau^+\tau^-$ yields, mis-modeling of MC samples, single candidate selection, the SCF ratio, signal and background modeling, fit bias, and interference.

We estimate the systematic uncertainty due to the resolution function by changing the resolution function parameters one by one by the uncertainties resulting from the $B^0\to D^{*-}\pi^{+}$ calibration procedure. For parameters that are not calibrated, we allow them to float in the fit and take the resulting shifts in $S$ and $C$ as the systematic uncertainties.

We validate the $\Delta t$ PDF shapes for $B \Bbar$ and \qqbar backgrounds by performing a fit to events in the $M_{\rm{bc}}$-sideband region.
The systematic uncertainty associated with these $\Delta t$ PDF shapes is estimated by varying the shape parameters in the $\Delta t$ PDF for $B \Bbar$ and $\qqbar$ backgrounds and repeating the fit.

The interference between CKM-favored and CKM-suppressed tag-side decays affects the values of $S$ and $C$ measured on the signal side~\cite{Long_2003}.
We generate simulated datasets both with and without interference effects and take the shifts in the values of $S$ and $C$ as the systematic uncertainties.

The wrong-tag fractions are calibrated as described in Ref.~\cite{belleiicollaboration2024new}. We estimate the uncertainty due to these wrong-tag fractions by varying the fractions by their uncertainties and repeating the fits. The resulting changes in the fit results from the nominal values are taken as the systematic uncertainties.

The values listed in Table~\ref{tab:peaking_cpv} allow for possible $CP$ violation in the peaking backgrounds. We generate simulated datasets varying the {\it CP} violating parameters for the backgrounds one at a time. The systematic uncertainty is estimated by taking the quadratic sum of the observed shifts.
We estimate the uncertainty due to possible {\it CP} violation in the TP signal in the same way, considering 50\% $CP$ violation in TP signal events and taking $f_L = 0.92$.

We estimate the uncertainty due to possible misalignment of the tracking detector \cite{Bilka:2020kgr}. We reconstruct a simulated sample for $B^0\to\rho^+\rho^-$ assuming four detector misalignment scenarios and extract $S$ and $C$. The systematic uncertainty is taken to be the maximum deviation from the nominal values.
We estimate the systematic uncertainty due to fixed physics parameters $\tau_{B^0}$ and $\Delta m_d$ by varying them by their uncertainties~\cite{ParticleDataGroup:2024xxx}.

\begin{table}[htbp]
\renewcommand{\arraystretch}{1.2}
  \centering
  \caption{Systematic uncertainties for $S$ and $C$.}
\begin{tabular*}{\linewidth}{@{\extracolsep{\fill}}lcc}
  \hline \hline
    Source & $S[10^{-2}]$&$C[10^{-2}]$\\
    \hline
$\mathcal{B}$'s of peaking backgrounds & $^{ +0.6 }_{-0.5}$ &$\pm 0.1$ \\
$\tau\tau$ background yield & $\pm 0.9$ &$^{ +0.0 }_{-0.1}$ \\
Data-MC mis-modeling & $^{ +0.6 }_{-1.1}$ &$^{ +1.5 }_{-0.6}$ \\
Single candidate selection & $\pm 1.3$ &$\pm 1.9$ \\
SCF ratio & $^{ +0.5 }_{-0.4}$ &$^{ +0.7 }_{-0.0}$ \\
Signal model & $^{ +1.1 }_{-1.4}$ &$^{ +0.3 }_{-0.4}$ \\
$q\bar{q}$ model & $^{ +2.2 }_{-1.0}$ &$\pm 0.2$ \\
$B\bar{B}$ model & $\pm 0.9$ &$^{ +0.7 }_{-0.5}$ \\
$\tau^+\tau^-$ model & $\pm 0.1$ &$\pm 0.0$ \\
Peaking model & $^{ +0.8 }_{-0.4}$ &$^{ +0.2 }_{-0.4}$ \\
Fit bias & $\pm 2.0$ &$\pm 0.6$ \\
Interference & $\pm 2.8$ &$\pm 1.7$ \\
Resolution & $^{ +3.4 }_{-4.4}$ &$^{ +1.9 }_{-1.4}$ \\
$\Delta t$ PDF for $q\bar{q}$ and $B\bar{B}$ & $^{ +3.8 }_{-1.8}$ &$^{ +0.7 }_{-0.1}$ \\
Tag side interference & $\pm 0.5$ &$\pm 2.1$ \\
Wrong tag fraction & $^{ +0.2 }_{-0.3}$ &$\pm 0.5$ \\
Background $CP$ violation & $^{ +3.8 }_{-3.6}$ &$^{ +4.2 }_{-3.7}$ \\
$CP$ violation in TP signal & $^{ +0.8 }_{-0.2}$ &$^{ +0.2 }_{-0.4}$ \\
Tracking detector misalignment & $\pm 1.4$ &$\pm 0.5$ \\
$\tau_{B^0}$ and $\Delta m_d$ & $^{ +1.4 }_{-1.6}$ &$\pm 0.3$ \\
\hline \hline
Total systematic uncertainty & $^{ +8.2 }_{-7.8}$ &$^{ +6.1 }_{-5.3}$ \\ 
Statistical uncertainty   &$\pm 18.8$ &{$\pm 12.1$} \\
\hline \hline
  \end{tabular*}
  \label{tab:total_uncertaintyCPV}
\end{table}

\subsection{Correlation}
Table~\ref{tab:correlation} summarizes the correlations among the four results for the statistical and systematic uncertainties.
We first estimate the correlations for each uncertainty individually. For example, we vary the PDF shape parameters by their uncertainties and record the changes for each pair of measurements to estimate the correlation of the modeling. This procedure is applied to all sources of systematics. 
The fitter gives the correlation of the statistical uncertainty between $\mathcal{B}$ and $f_L$, as well as between $S$ and $C$. The statistical correlation between $\mathcal{B}$ or $f_L$ and $S$ or $C$ is estimated by varying $\mathcal{B}$ or $f_L$ by $1\sigma$ and then repeating the {\it CP} fit.

 \begin{table}[htbp]
 \label{tab:correlation}
  \centering
  \caption{
  The correlations among the measurements for the statistical and systematic uncertainties. 
  }
  \begin{tabular}{l|rrrr}
  \hline \hline
  Stat.       & $f_{L}$ & $S$ & $ C$\\
\hline 
$\mathcal{B}$ & $-0.11$ & 0.04 & $0.01$ \\
$f_{L}$       &         & 0.03 & $0.05$ \\
$S$           &         &      & $-0.06$ \\
    \hline \hline
  Syst.       & $f_{L}$ & $S$ & $ C$\\
\hline
$\mathcal{B}$ & $-0.12$ & 0.06    & 0.01 \\
$f_{L}$       &         & $-0.03$ & 0.00 \\
$S$           &         &         & 0.02 \\
  \hline \hline
  \end{tabular}
  \label{tab:correlation}
\end{table}

\section{Constraints on the CKM angle $\phi_2$}
\label{sec:phi2}
We extract $\phi_2$ by performing an isospin analysis using the method in Ref.~\cite{Belle:2015xfb} based on the Gronau-London isospin relations~\cite{Gronau:1990ka}
\begin{align}
  \frac{1}{\sqrt{2}} A_{+-} + A_{00} &= A_{+0},\\
  \frac{1}{\sqrt{2}} \bar{A}_{+-} + \bar{A}_{00} &= \bar{A}_{-0},
\end{align}
where $A_{ij}$ is the amplitude of longitudinally polarized $B\to\rho^i\rho^j$.
We use results from the Belle, \babar, and LHCb experiments for $\mathcal{B}$, $f_L$, and {\it CP} violation parameters for $B^0\to\rho^+\rho^-$, $B^0\to\rho^0\rho^0$, and $B^+\to\rho^+\rho^0$ decays, and the ratio of $B^+$ and $B^0$ lifetimes as listed in the PDG~\cite{ParticleDataGroup:2024xxx}. 
Since Belle and \babar assumed equal production of $B^+B^-$ and $B^0\Bbar^0$ pairs, we correct their $\cal{B}$ values to account for the latest HFLAV value of $f_{+-}$ 
and $f_{00}$~\cite{banerjee2024averagesbhadronchadrontaulepton}.
The inclusion of this effect slightly increases the branching fractions for the $B^0$ mode and slightly decreases that for the $B^+$ mode. The result of the isospin analysis is $\phi_2$ = $(91.5^{+4.8}_{-5.2})^\circ$ and $\Delta \phi_2=(2.4^{+4.2}_{-3.8})^\circ$. The updated values of $f_{+-}$ and $f_{00}$ shift $\phi_2$ by $-0.4^\circ$.

\begin{figure}[htb]
  \centering
  \includegraphics[width=0.45\textwidth]{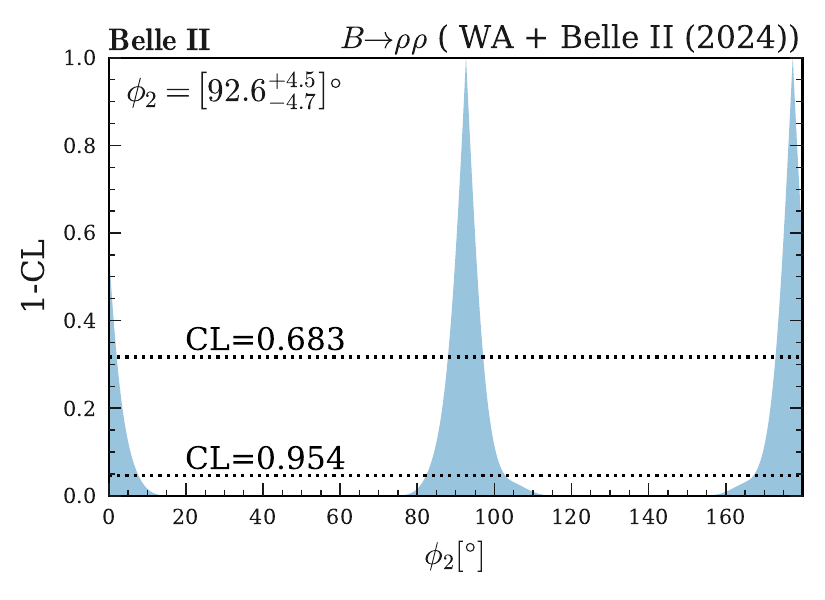}
  \caption{Probability ($1-$Confidence-Level) for the CKM angle $\phi_2$ based on combined inputs from the world averages~\cite{ParticleDataGroup:2024xxx} and our results of $B\to\rho\rho$ decays. The black dotted lines correspond to the 0.683 and 0.954 confidence levels.
  }
  \label{fig:phi2}
\end{figure}
We subsequently combine our $B^0\to\rho^+\rho^-$ results with other results and extract $\phi_2$. The results are $\phi_2 = (92.6^{+4.5}_{-4.7})^\circ$ and $\Delta \phi_2 = (2.4^{+3.8}_{-3.7} )^\circ$. The likelihood curve is shown in Figure~\ref{fig:phi2}. Our isospin analysis yields a second solution of $\phi_2 = (177.4 ^{+4.7}_{-4.5})^\circ$ and $\Delta \phi_2 = (-2.4^{+3.7}_{-3.8} )^\circ$; however, this value for $\phi_2$ is excluded by measurements of the UT angles $\phi_1$ and $\phi_3$~\cite{ParticleDataGroup:2024xxx} and unitarity.
The dominant uncertainties on $\phi_2$ are due to the $S$ parameters for $B^0 \to \rho^+ \rho^-$ and $B^0\to\rho^0\rho^0$.

\section{Conclusions}
\label{sec:conclusions}
We have measured the branching fraction and longitudinal polarization fraction for $B^0 \to \rho^+ \rho^-$ decays as well as {\it CP} violation parameters for the longitudinal polarized decay using a data sample of $(387\pm6) \times 10^6$ $\Upsilon(4S)$ decays.
We obtain
\begin{align}
  \mathcal{B}(B^0\to\rho^+\rho^-) &= \left(2.89 ^{+0.23}_{-0.22} {}^{+0.29}_{-0.27}\right) \times 10^{-5},\\
  f_{L} &= 0.921 ^{+0.024}_{-0.025} {}^{+0.017}_{-0.015},\\
    S &= -0.26\pm0.19\pm0.08,\\
  C &= -0.02\pm0.12^{+0.06}_{-0.05},
\end{align}
where the first uncertainties are statistical and the second are systematic. The value for $f_L$ is somewhat lower than the world average but consistent with it within $2\sigma$.
These results are in good agreement with previous measurements~\cite{Belle:2015xfb,BaBar:2007cku}.
We constrain $\phi_2$ using our results as well as $B \to \rho\rho$ results from other experiments, adjusting the latter to account for the most recent values of $f_{+-}$ and $f_{00}$~\cite{banerjee2024averagesbhadronchadrontaulepton}.
We obtain $\phi_2 = (92.6^{+4.5}_{-4.7})^\circ$, which is consistent with other UT observables.
The uncertainty is dominated by the precision of the $S$ parameters for $B^0 \to \rho^+ \rho^-$ and $B^0 \to \rho^0\rho^0$, which can be improved by future measurements.
This result can be used to constrain non-SM physics.

\section*{Acknowledgements}
This work, based on data collected using the Belle II detector, which was built and commissioned prior to March 2019,
was supported by
Higher Education and Science Committee of the Republic of Armenia Grant No.~23LCG-1C011;
Australian Research Council and Research Grants
No.~DP200101792, 
No.~DP210101900, 
No.~DP210102831, 
No.~DE220100462, 
No.~LE210100098, 
and
No.~LE230100085; 
Austrian Federal Ministry of Education, Science and Research,
Austrian Science Fund
No.~P~34529,
No.~J~4731,
No.~J~4625,
and
No.~M~3153,
and
Horizon 2020 ERC Starting Grant No.~947006 ``InterLeptons'';
Natural Sciences and Engineering Research Council of Canada, Compute Canada and CANARIE;
National Key R\&D Program of China under Contract No.~2022YFA1601903,
National Natural Science Foundation of China and Research Grants
No.~11575017,
No.~11761141009,
No.~11705209,
No.~11975076,
No.~12135005,
No.~12150004,
No.~12161141008,
No.~12475093,
and
No.~12175041,
and Shandong Provincial Natural Science Foundation Project~ZR2022JQ02;
the Czech Science Foundation Grant No.~22-18469S 
and
Charles University Grant Agency project No.~246122;
European Research Council, Seventh Framework PIEF-GA-2013-622527,
Horizon 2020 ERC-Advanced Grants No.~267104 and No.~884719,
Horizon 2020 ERC-Consolidator Grant No.~819127,
Horizon 2020 Marie Sklodowska-Curie Grant Agreement No.~700525 ``NIOBE''
and
No.~101026516,
and
Horizon 2020 Marie Sklodowska-Curie RISE project JENNIFER2 Grant Agreement No.~822070 (European grants);
L'Institut National de Physique Nucl\'{e}aire et de Physique des Particules (IN2P3) du CNRS
and
L'Agence Nationale de la Recherche (ANR) under grant ANR-21-CE31-0009 (France);
BMBF, DFG, HGF, MPG, and AvH Foundation (Germany);
Department of Atomic Energy under Project Identification No.~RTI 4002,
Department of Science and Technology,
and
UPES SEED funding programs
No.~UPES/R\&D-SEED-INFRA/17052023/01 and
No.~UPES/R\&D-SOE/20062022/06 (India);
Israel Science Foundation Grant No.~2476/17,
U.S.-Israel Binational Science Foundation Grant No.~2016113, and
Israel Ministry of Science Grant No.~3-16543;
Istituto Nazionale di Fisica Nucleare and the Research Grants BELLE2;
Japan Society for the Promotion of Science, Grant-in-Aid for Scientific Research Grants
No.~16H03968,
No.~16H03993,
No.~16H06492,
No.~16K05323,
No.~17H01133,
No.~17H05405,
No.~18K03621,
No.~18H03710,
No.~18H05226,
No.~19H00682, 
No.~20H05850,
No.~20H05858,
No.~22H00144,
No.~22K14056,
No.~22K21347,
No.~23H05433,
No.~26220706,
and
No.~26400255,
and
the Ministry of Education, Culture, Sports, Science, and Technology (MEXT) of Japan;  
National Research Foundation (NRF) of Korea Grants
No.~2016R1-D1A1B-02012900,
No.~2018R1-A6A1A-06024970,
No.~2021R1-A6A1A-03043957,
No.~2021R1-F1A-1060423,
No.~2021R1-F1A-1064008,
No.~2022R1-A2C-1003993,
No.~2022R1-A2C-1092335,
No.~RS-2023-00208693,
No.~RS-2024-00354342
and
No.~RS-2022-00197659,
Radiation Science Research Institute,
Foreign Large-Size Research Facility Application Supporting project,
the Global Science Experimental Data Hub Center, the Korea Institute of
Science and Technology Information (K24L2M1C4)
and
KREONET/GLORIAD;
Universiti Malaya RU grant, Akademi Sains Malaysia, and Ministry of Education Malaysia;
Frontiers of Science Program Contracts
No.~FOINS-296,
No.~CB-221329,
No.~CB-236394,
No.~CB-254409,
and
No.~CB-180023, and SEP-CINVESTAV Research Grant No.~237 (Mexico);
the Polish Ministry of Science and Higher Education and the National Science Center;
the Ministry of Science and Higher Education of the Russian Federation
and
the HSE University Basic Research Program, Moscow;
University of Tabuk Research Grants
No.~S-0256-1438 and No.~S-0280-1439 (Saudi Arabia), and
King Saud University,Riyadh, Researchers Supporting Project number (RSPD2024R873)  
(Saudi Arabia);
Slovenian Research Agency and Research Grants
No.~J1-9124
and
No.~P1-0135;
Agencia Estatal de Investigacion, Spain
Grant No.~RYC2020-029875-I
and
Generalitat Valenciana, Spain
Grant No.~CIDEGENT/2018/020;
The Knut and Alice Wallenberg Foundation (Sweden), Contracts No.~2021.0174 and No.~2021.0299;
National Science and Technology Council,
and
Ministry of Education (Taiwan);
Thailand Center of Excellence in Physics;
TUBITAK ULAKBIM (Turkey);
National Research Foundation of Ukraine, Project No.~2020.02/0257,
and
Ministry of Education and Science of Ukraine;
the U.S. National Science Foundation and Research Grants
No.~PHY-1913789 
and
No.~PHY-2111604, 
and the U.S. Department of Energy and Research Awards
No.~DE-AC06-76RLO1830, 
No.~DE-SC0007983, 
No.~DE-SC0009824, 
No.~DE-SC0009973, 
No.~DE-SC0010007, 
No.~DE-SC0010073, 
No.~DE-SC0010118, 
No.~DE-SC0010504, 
No.~DE-SC0011784, 
No.~DE-SC0012704, 
No.~DE-SC0019230, 
No.~DE-SC0021274, 
No.~DE-SC0021616, 
No.~DE-SC0022350, 
No.~DE-SC0023470; 
and
the Vietnam Academy of Science and Technology (VAST) under Grants
No.~NVCC.05.12/22-23
and
No.~DL0000.02/24-25.

These acknowledgements are not to be interpreted as an endorsement of any statement made
by any of our institutes, funding agencies, governments, or their representatives.

We thank the SuperKEKB team for delivering high-luminosity collisions;
the KEK cryogenics group for the efficient operation of the detector solenoid magnet and IBBelle on site;
the KEK Computer Research Center for on-site computing support; the NII for SINET6 network support;
and the raw-data centers hosted by BNL, DESY, GridKa, IN2P3, INFN, 
and the University of Victoria.

\bibliographystyle{apsrev4-1}
\bibliography{references}

\begin{thebibliography}{55}
\bibitem [{Notea)()}]{Notea)}%
  \BibitemOpen
  \bibinfo {note} {The angle $\phi _2$ is also known as $\alpha $.}\BibitemShut
  {Stop}%
\bibitem [{Noteb)()}]{Noteb)}%
  \BibitemOpen
  \bibinfo {note} {We use a system of units in which $c=\hbar =1$.}\BibitemShut
  {Stop}%
\bibitem [{Notee)()}]{Notee)}%
  \BibitemOpen
  \bibinfo {note} {Throughout this paper, the inclusion of the charge conjugate
  decay mode is implied unless otherwise specified.}\BibitemShut {Stop}%
\bibitem [{Notef)()}]{Notef)}%
  \BibitemOpen
  \bibinfo {note} {A small number of non-resonant $e^+e^-\protect \ensuremath
  {\rightarrow }\protect \xspace B\protect \bar {B}$ events are also included
  in the number of $Y(4S)$ decays; these events are indistinguishable from
  $e^+e^-\protect \ensuremath {\rightarrow }\protect \xspace \Upsilon
  (4S)\protect \ensuremath {\rightarrow }\protect \xspace B\protect \bar
  {B}$.}\BibitemShut {Stop}%
\end{thebibliography}

\end{document}